\documentclass[sigplan,10pt]{acmart}

\acmSubmissionID{526} 

\AtBeginDocument{%
	}

\copyrightyear{2024}
\acmYear{2024}
\setcopyright{rightsretained}
\acmConference[EuroSys '24]{Nineteenth European Conference on Computer Systems}{April 22--25, 2024}{Athens, Greece}
\acmBooktitle{Nineteenth European Conference on Computer Systems (EuroSys '24), April 22--25, 2024, Athens, Greece}\acmDOI{10.1145/3627703.3629587}
\acmISBN{979-8-4007-0437-6/24/04}

\usepackage{lipsum}  
\usepackage{subcaption} 
\usepackage{enumitem}
\usepackage{amsmath}
\usepackage{bbm}

\DeclareMathOperator*{\argmin}{argmin}

\begin{document}
\newcommand{\scheme}{Atlas}

\title{\scheme{}: Hybrid Cloud Migration Advisor for\\Interactive Microservices}

\renewcommand{\shorttitle}{\scheme{}: Hybrid Cloud Migration Advisor for Interactive Microservices}

\author{Ka-Ho Chow$^{+*}$, Umesh Deshpande$^{*}$, Veera Deenadayalan$^{*}$, Sangeetha Seshadri$^{*}$, Ling Liu$^{+}$}
\affiliation{%
	\institution{$^{+}$ Georgia Institute of Technology, Atlanta, Georgia, USA}
	\institution{$^{*}$ IBM Research - Almaden, San Jose, California, USA}
	\country{}
}
\renewcommand{\authors}{Ka-Ho Chow, Umesh Deshpande, Veera Deenadayalan, Sangeetha Seshadri, and Ling Liu}
\renewcommand{\shortauthors}{Ka-Ho Chow, Umesh Deshpande, Veera Deenadayalan, Sangeetha Seshadri, and Ling Liu}

\begin{abstract}
Hybrid cloud provides an attractive solution to microservices for better resource elasticity. A subset of application components can be offloaded from the on-premises cluster to the cloud, where they can readily access additional resources. However, the selection of this subset is challenging because of the large number of possible combinations. A poor choice degrades the application performance, disrupts the critical services, and increases the cost to the extent of making the use of hybrid cloud unviable. This paper presents \scheme{}, a hybrid cloud migration advisor. \scheme{} uses a data-driven approach to learn how each user-facing API utilizes different components and their network footprints to drive the migration decision. It learns to accelerate the discovery of high-quality migration plans from millions and offers recommendations with customizable trade-offs among three quality indicators: end-to-end latency of user-facing APIs representing application performance, service availability, and cloud hosting costs. \scheme{} continuously monitors the application even after the migration for proactive recommendations. Our evaluation shows that \scheme{} can achieve $21\%$ better API performance (latency) and $11\%$ cheaper cost with less service disruption than widely used solutions. 
\end{abstract}


	
\maketitle
\section{Introduction}
\begin{figure}
	\begin{minipage}{0.985\linewidth}
		\centering
		\includegraphics[width=\linewidth]{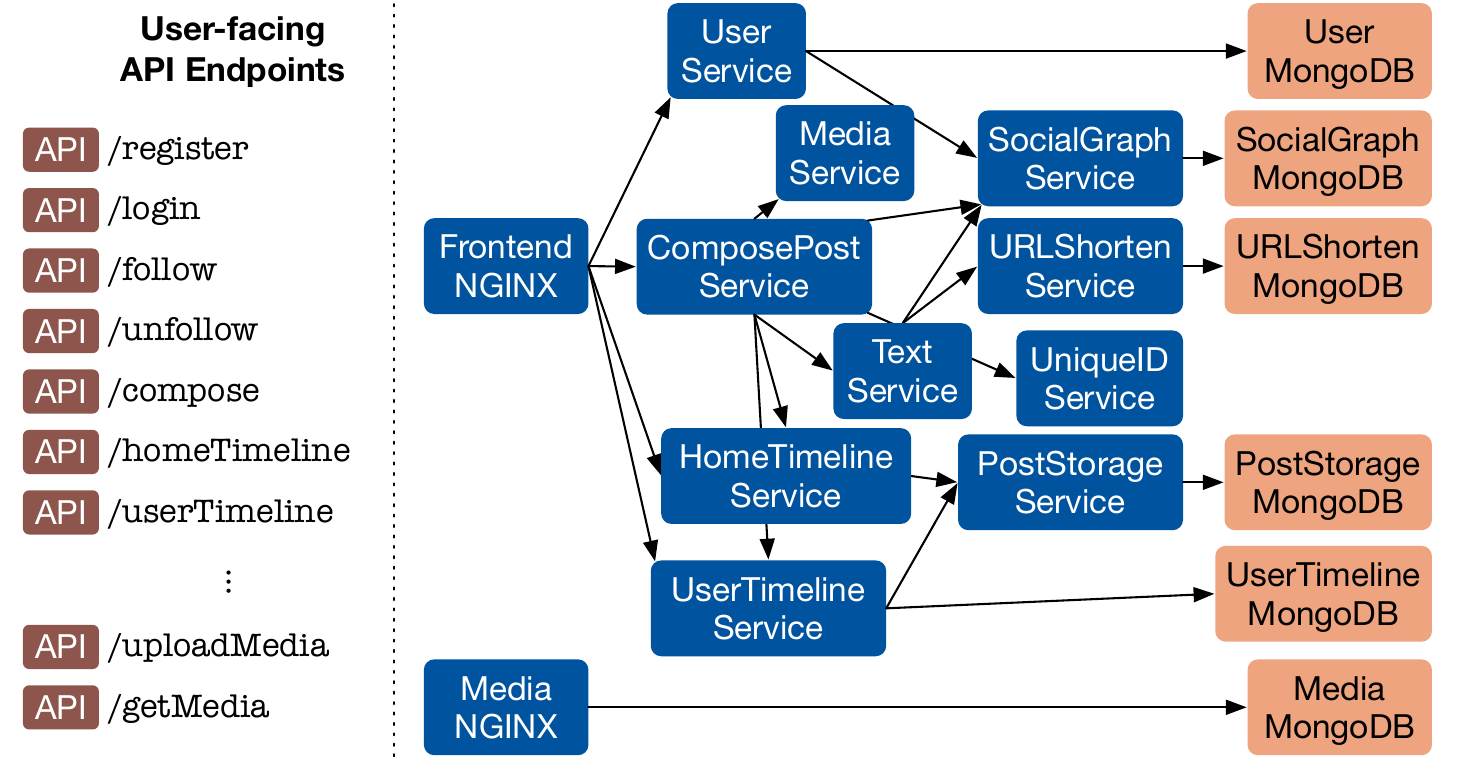}\vspace{-0.5em}
		\caption{The nine user-facing APIs and their simplified flows across components in a social network from DeathStarBench~\cite{gan2019open}.}
		\label{sn-arch}
	\end{minipage}\vspace{0.5em}
	\begin{minipage}{0.99\linewidth}
		\centering
		\includegraphics[width=\linewidth]{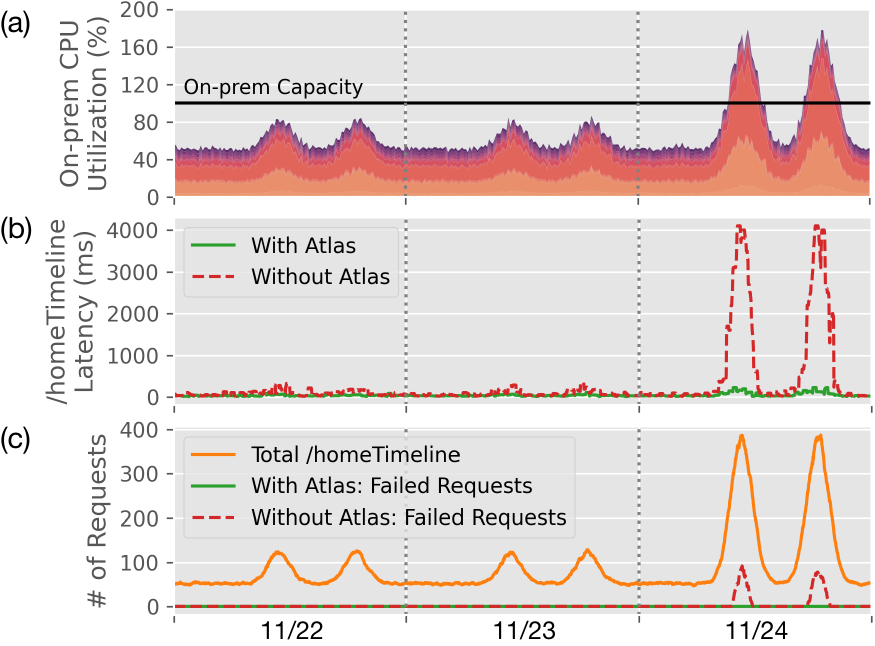}\vspace{-0.5em}
		\caption{API request latency spikes and failures due to inelastic infrastructure to handle user requests on the social network (Figure~\ref{sn-arch}). }
		\label{fig:intro-overload}
	\end{minipage}
\end{figure}
Hybrid cloud enables applications to seamlessly access both on-premises (on-prem) and public cloud infrastructure. While on-prem infrastructure offers greater security and control over data, public clouds excel in scalability and lower capital expenditure~\cite{netapp}. Microservices are a major driver behind the hybrid cloud since they allow an application to be broken into smaller, purpose-built components, which can be placed on the infrastructure that benefits them the most~\cite{ibm-hybridcloud}. 

Taking a social network as an example in Figure~\ref{sn-arch}, a burst in user requests can lead to the CPU demand exceeding the on-prem capacity (Figure~\ref{fig:intro-overload}a). 
It can be Thanksgiving on 11/24 or, in general, any planned burst, such as seasonal behaviors. Such an overloaded application slows down the responsiveness of API requests (e.g., $\texttt{/homeTimeline}$ in Figure~\ref{fig:intro-overload}b) or even leads to request failures in Figure~\ref{fig:intro-overload}c. Purchasing additional resources for the on-prem cluster is not economical because of the expected drop in utilization after the peak season. Completely migrating to the cloud may not be feasible as well because of privacy regulations limiting the relocation of business-critical data~\cite{EUdataregulations2018}. In this context, hybrid cloud enables the application to offload some components to the cloud, where they can readily access additional resources. 
Additionally, hybrid cloud can benefit microservices in handling application growth~\cite{gilgur2015precentile}, lowering resource consumption to accommodate other applications hosted in the same cluster~\cite{sadashiv2011cluster}, or reducing the carbon footprint for sustainable computing~\cite{greencomputing}.

\begin{figure}
	\centering
	\includegraphics[width=\linewidth]{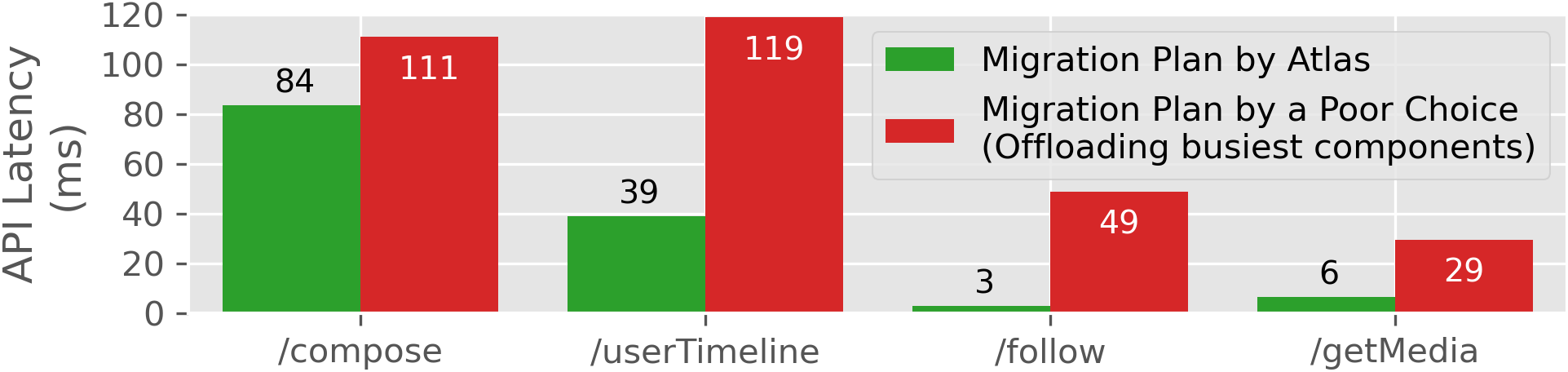}\vspace{-0.5em}
	\caption{A poor choice of components to offload can lead to over $10\times$ more performance degradation than our recommendation.}\label{fig:intro-preview}\vspace{-0.8em}
\end{figure}

The use of hybrid clouds may lead to several concerns, especially when a poor choice of components is offloaded. Application components communicate with each other to jointly offer services through user-facing APIs. When some are offloaded, inter-datacenter communications may be required to serve an API request. A component triggering another one on a different location must wait for a longer network-induced latency. Those inter-datacenter communications collectively prolong the end-to-end latency for an API request and impact user experience (see the red bars in Figure~\ref{fig:intro-preview}). 
Furthermore, the availability of APIs can also be affected because some may suffer from downtime during the migration process (e.g., data transfer for stateful components to the cloud). Last, cloud hosting costs can be unnecessarily high for underutilized cloud nodes and egress traffic, which can be avoided by intelligently placing components.

Selecting a suitable subset of components to offload while alleviating the above downsides is challenging. For instance, the social network in Figure~\ref{sn-arch} with $29$ components leads to over $500$ million combinations. Several strategies have been developed to make such a decision more systematic: (i) the component-focused solutions to offload the most resource-consuming ones~\cite{guo2012seagull} and (ii) the affinity-based solutions to minimize the traffic between on-prem and cloud datacenters~\cite{sampaio2019improving,joseph2020intma,hu2019optimizing,han2020refining}. However, they do not consider how components are being used end-to-end to serve different API requests but only focus on individual component usages or how two components interact in general. User-facing APIs are of primary concern to both application owners and their users because they often reflect business logic (e.g., a \texttt{/purchase} API request in an online store refers to a consumer purchasing a product). Some APIs are more critical than others (e.g., \texttt{/purchase} versus \texttt{/bookmark}). Maintaining their responsiveness and availability is necessary as they directly impact user experience and business revenue~\cite{latencyreduction}.

Despite its benefits, finding migration plans with an API-centric  design is challenging for several reasons. First, it is impractical to compare plans by conducting actual migration to observe the API latency, service disruption, and cost. Second, a component can be used by multiple APIs and triggered with different workflows. Offloading a component can lead to significant performance degradation in some APIs due to long-distance communications but can also have little to no impact on other APIs using it. Third, high-quality plans should be pinpointed from a search space that grows exponentially with the number of application components. This paper addresses the above challenges by proposing \scheme{}, a hybrid cloud migration advisor. \scheme{} uses a data-driven approach to learn the characteristics of each user-facing API and ensures their quality in terms of latency and availability after migration. Our main contributions are as follows:
\begin{itemize}[leftmargin=*,noitemsep,topsep=0pt]
	\item We propose a migration plan recommender system with user-facing APIs as first-class citizens, optimizing API latency, API availability, and cloud hosting cost. It adapts to the business needs of the application owner and monitors the application for proactive recommendations.
	\item We develop a suite of techniques to characterize each user-facing API, including its execution workflow and network footprint, from telemetry data. This allows \scheme{} to (i) recommend components based on how they are being used in serving requests of different APIs and (ii) be deployed for any application without modification.
	\item We introduce a deep reinforcement learning-based genetic algorithm for accelerated recommendations, which avoids low-quality and infeasible migration plans.
\end{itemize}
We use \scheme{} to advise migration for a \emph{social network} and a \emph{hotel reservation system} from DeathStarBench~\cite{gan2019open} to demonstrate its high-quality migration plans and customizable trade-offs. We envision that \scheme{} can be deployed as a service and is a building block for sky computing~\cite{sky-ibm,stoica2021cloud}.

\section{Related Work}
When resource usage is expected to exceed capacity, some workloads can be offloaded to the cloud. The busiest components are often relocated first as they can free up the most resources~\cite{guo2012seagull}. This component-focused policy is the easiest to implement to automate the bursting~\cite{aws-cb,azure-cb}, but it does not consider the impact of migration on the application.

Advanced microservice placement approaches assign components to nodes (or clusters) based on certain quality indicators~\cite{aksakalli2021deployment,zhong2022machine}. RL-MA~\cite{joseph2019fuzzy} finds a placement plan with minimal energy consumption using the least number of nodes but does not consider the impact on application performance due to inter-node communications. REMaP~\cite{sampaio2019improving}, IntMA~\cite{joseph2020intma}, \citet{hu2019optimizing}, and \citet{han2020refining} manage the placement by minimizing the affinity between components in different nodes, considering the overall traffic size and/or the number of message exchanges over time. \citet{aznavouridis2022micro} and \citet{ding2022kubernetes} use a cost-driven approach to find placement plans to minimize traffic and compute costs. \citet{guerrero2018resource} search for placement plans that minimize the overall network delay, the compute cost, and the time to download the Docker image to a node. These approaches minimize the traffic between nodes by tracking the communication between different components. However, they do not consider the contribution of a given component in serving an end-to-end API request. We show that the API-centric view creates unique opportunities to reduce performance degradation and cost. For instance, offloading a component could incur additional traffic but with little to no impact on the end-to-end latency of APIs using it. \citet{bhamare2017multi} place services across clouds, however, only with explicit knowledge of the application's workflow w.r.t. each API.

\scheme{} is unique in the following aspects compared to the above intra-cluster and inter-cloud solutions. First, \scheme{} does not assume any knowledge of the application logic, and it only uses the readily available telemetry data to optimize end-to-end API latency. Second, \scheme{} incorporates the application owner's preferences to weigh the migration plans and minimize the disruption of critical APIs. Finally, \scheme{} considers the cloud's cost model of dynamic resource allocation (autoscaling) and recommends migrating components that could best leverage autoscaling to reduce cost.

\section{\scheme{} Overview}\label{sec:bg}
\scheme{} is a hybrid cloud migration advisor designed as a loop of three stages: (i) application learning, (ii) migration recommendation, and (iii) post-migration monitoring for proactive recommendations. It targets API-driven interactive microservices. These applications offer services by exposing user-facing API endpoints, and their clients can invoke them through, e.g., HTTP requests. Each API endpoint is often implemented for a specific task and can require specific inputs as a payload from the client. Once an API request reaches the entry component (e.g., \texttt{FrontendNGINX}), it is processed based on the implemented business logic, and other components may be triggered to jointly complete the task before sending the response to the client. An interactive application is latency-sensitive. It demands an API request be served instantly, and hence the end-to-end latency of each API endpoint should be minimized to ensure responsiveness.

\begin{figure}
	\centering
	\includegraphics[width=\linewidth]{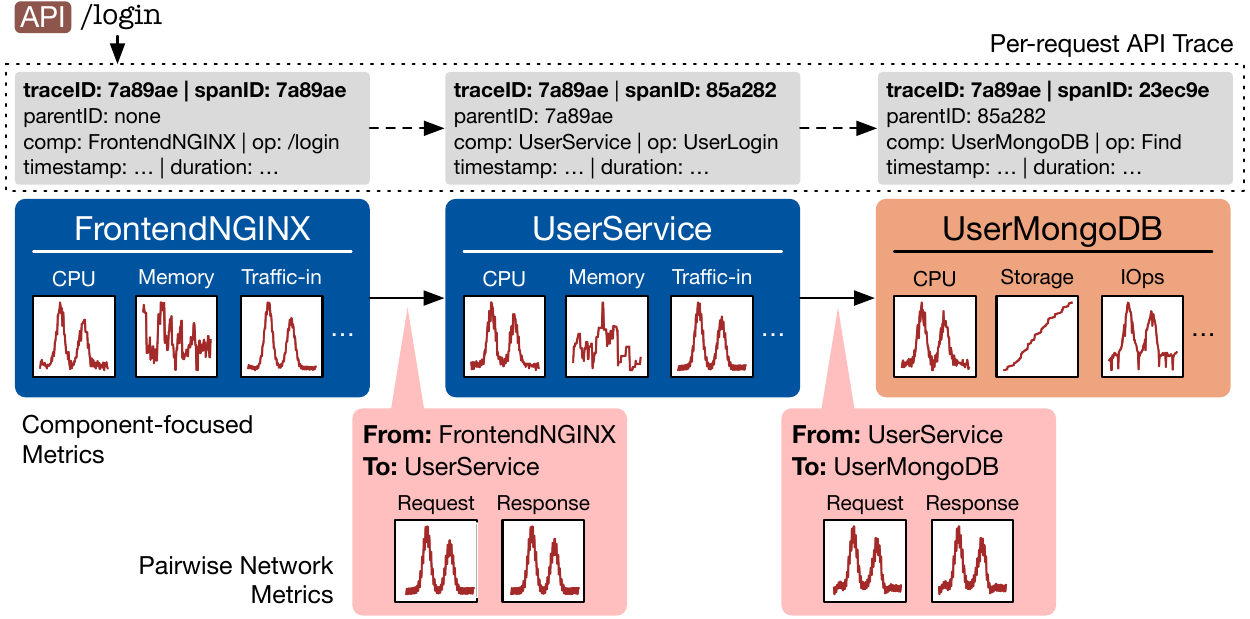}
	\caption{\scheme{} is purely driven by telemetry data: component-focused metrics, network metrics, and per-request API traces.}\label{fig:telemetry}
\end{figure}

\textbf{Observability-driven Advisor.} \scheme{} is a data-driven solution fueled by three types of telemetry data to learn the application before making decisions. First, component-focused metrics, often from cAdvisor~\cite{cadvisor}, record resource consumption such as CPU, memory, ingress, and egress traffic of each component (container) over time, as shown in Figure~\ref{fig:telemetry} with three example components. 
Second, more fine-grained network metrics can be observed by Istio~\cite{istio}, which records how many bytes are being transferred from one component to another during the request and the response over time (see the pink boxes). Third, distributed tracing~\cite{sigelman2010dapper} provides the signals for \scheme{} to learn the application logic of APIs. For every API request the application receives, the tracing library (e.g., Jaeger~\cite{jaeger}) creates an instance of a data structure called trace with a unique ID. A trace consists of several spans, and each corresponds to an operation done by the application to serve the request. Taking the \texttt{/login} API in Figure~\ref{fig:telemetry} as an example, when the request from the client reaches the entry component \texttt{FrontendNGINX}, a root span (the first gray box) is created with attributes such as the component name, the operation name, the timestamp it is triggered, and the duration. Then, it invokes \texttt{UserService}, represented by a child span (the middle gray box), which finally consults \texttt{UserMongoDB} creating the third span. Each child span includes the ID of the parent initiating it, and hence a trace encapsulates the entire lifetime of an API request in the application. Harvesting knowledge from it allows \scheme{} to understand API logic in a data-driven manner.

\begin{figure*}
	\centering
	\includegraphics[width=\linewidth]{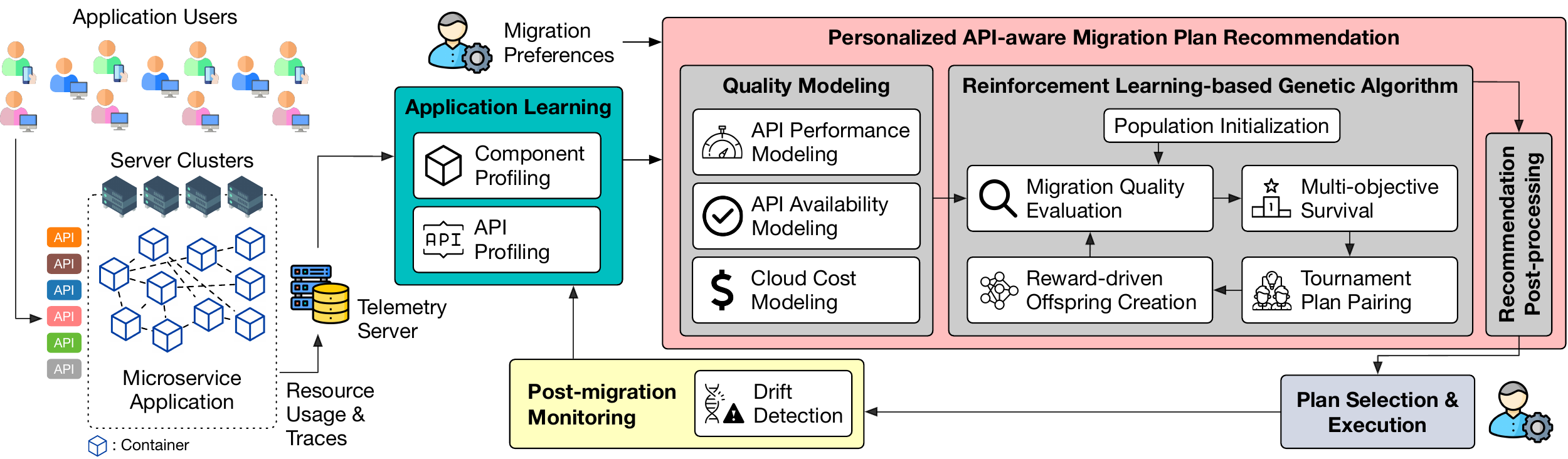}
	\caption{\scheme{} is a loop of three stages with (i) application learning, (ii) migration recommendation, and (iii) post-migration monitoring.}\label{fig:overview}
\end{figure*}
\textbf{Elastic Microservices.} \scheme{} aims to optimize the use of public clouds to extend the elasticity of on-prem infrastructure. Cloud providers offer virtually unlimited resources with flexible pricing to only charge for consumed resources. For example, application owners can create a Kubernetes cluster~\cite{k8s} and use the built-in cluster autoscaler to automatically adjust the number of nodes being used based on resource consumption. This scaling operation can be done in minute-level granularity and hence is an economical approach to achieving better resource elasticity. Given a migration plan from \scheme{} specifying the placement of each component, the offloading can be done in various ways, depending on the cluster configuration. In a Kubernetes cluster with nodes at both on-prem and cloud datacenters, offloading a component is as simple as pointing the location to another address. Several tools have also been developed to ease relocation for clusters~\cite{vkubelet,crossplane,terraform}, catering to aspects such as security~\cite{submariner} and data transfer for stateful components~\cite{velero}.

\textbf{Migration Considerations.} The migration process can incur overheads to the computing infrastructure. For instance, \scheme{} assumes that stateful components can be relocated using data transfer tools like Velero~\cite{velero}. Moving them may consume network resources, and the exact overhead depends on the specific mechanism used for data transfer, e.g., continuous replication, and needs to be considered accordingly. \scheme{} focuses on the performance implication \emph{after} the migration. The application owners can determine the viability of the stateful components for migration and provide the preferences to \scheme{} for customized recommendations. For instance, when the migration is performed to serve a temporary burst, the movement of certain stateful components can be disabled to avoid excessive overhead.

\textbf{Design Principles.} \scheme{} is designed with the following principles for a broader impact and offering as a service.
\begin{itemize}[leftmargin=*,noitemsep,topsep=0pt]
	\item \emph{Non-intrusive and Easy-to-deploy}: Any application desired to enable \scheme{} only needs to include the libraries for monitoring and tracing, and they have become a standard today in microservice frameworks~\cite{opentelemetry}.
	\item \emph{Unsupervised Learning}: \scheme{} should not (i) assume any implementation knowledge of application components or APIs (e.g., call graphs) and (ii) require any custom workloads for supervision. Learning signals can only be derived from the information available in the production system.
	\item \emph{Privacy-preserving}: \scheme{} should rely on non-sensitive resource metrics and distributed traces without requiring any high-level information (e.g., logs) that can expose the application semantics. 
\end{itemize}

\textbf{\scheme{} Design.} We give an overview of \scheme{} in Figure~\ref{fig:overview}. During the application learning phase, \scheme{} queries the telemetry server to get the resource metrics and distributed traces collected from the production system to learn a profile of each component and each user-facing API. Then, the application owner provides migration preferences, including which APIs are critical to their business, budget, placement constraints (e.g., due to regulatory restrictions), resource limits, and the expected resource usage derived from historical patterns or any estimator~\cite{desnoyers2012modellus,verma2016dynamic,meng2016crupa,tran2018multivariate,yan2021hansel,zhou2021learning,chow2022deeprest,chow2023scad} within the period of interest (e.g., 11/24 in Figure~\ref{fig:intro-overload}). The migration recommendation module offers a list of migration plans optimizing API performance, API availability, and cloud hosting cost with different trade-offs. To simplify the plan selection experience, we use a hierarchical approach to pinpoint the best migration plan among several possible candidates. Finally, the application owner executes the selected plan, which triggers the post-migration monitoring stage to monitor the application status and trigger a new round of recommendations if better plans can be offered. 

\section{\scheme{} Methodology}
\subsection{Migration Quality Modeling}\label{sec:modeling}
Being able to compare the quality of migration plans is a prerequisite. \scheme{} models the quality without conducting actual migration and measurements, which are time-consuming and impractical. Let $\boldsymbol{p}$ be a migration plan for a microservice application with a set $\boldsymbol{\mathcal{C}}$ of components, and $p_c$ denotes the assigned location of component $c\in\boldsymbol{\mathcal{C}}$. \scheme{} supports multi-clouds, but for brevity, we focus the discussion on two locations: (i) on-prem with $p_c=0$ and (ii) cloud with $p_c=1$. 

\subsubsection{API Performance Modeling}
\begin{figure*}
	\centering
	\begin{minipage}[b]{.69\textwidth}
		\centering
		\includegraphics[width=\linewidth]{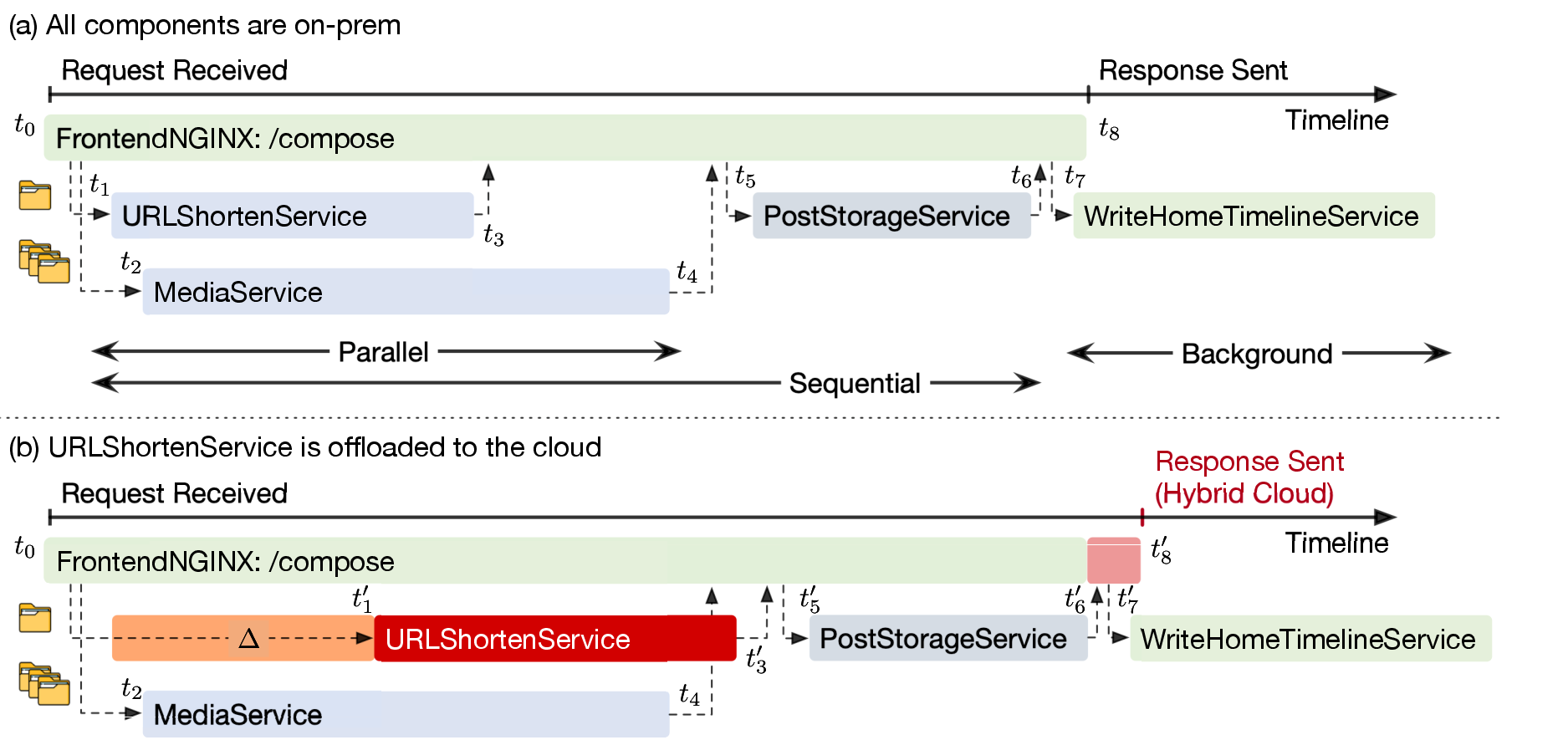}
		\caption{\scheme{} uses delay injection to estimate the API latency given a migration plan.}
		\label{fig:delay-injection}
	\end{minipage}\hfill
	\begin{minipage}[b]{.29\textwidth}
		\centering
		\includegraphics[width=\linewidth]{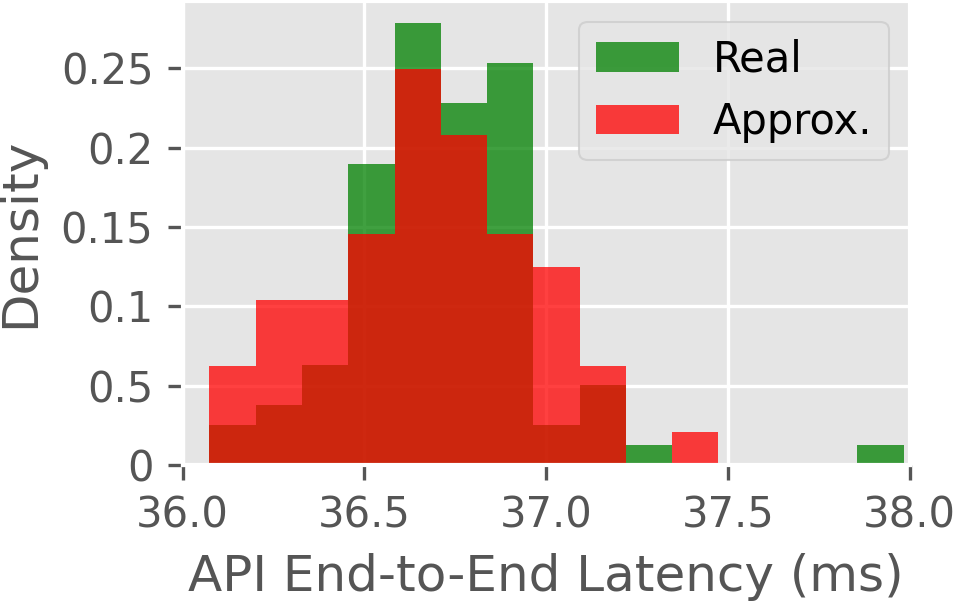}\vspace{-0.5em}
		\caption{For each user-facing API, we repeat the delay injection process on multiple existing traces to approximate its end-to-end latency distribution after the migration (red), which is close to the real one (green) collected for evaluation purposes.}\label{fig:footprint-dist0}
	\end{minipage}
\end{figure*}
\scheme{} optimizes the end-to-end latency of user-facing APIs because of its direct interest to both application owners and their clients. Such an approach allows \scheme{} to take advantage of the execution workflow of components in serving an API request to find the migration plan with minimal or even no impact on API responsiveness. We introduce a delay injection technique on \emph{collected traces} to estimate the latency for each user-facing API. It provides a preview of how APIs would perform after executing a migration plan to the application owner. It requires overcoming the challenges of (i) identifying where to inject the delay and its cascading effects on downstream operations and (ii) how much should be injected considering the network performance and the data size to be transferred.

\textbf{Execution Workflow.} Figure~\ref{fig:delay-injection}a shows the execution diagram of a simplified trace from a \texttt{/compose} request with carefully marked timestamps for illustration. This is a common visualization used by most tracing libraries~\cite{jaeger}. We use this trace to explain delay injection because it covers the patterns that are generalizable to any other user-facing APIs. The request first triggers \texttt{FrontendNGINX} at $t_0$, which further invokes both \texttt{URLShortenService} at $t_1$ and \texttt{MediaService} at $t_2$ to process the post content. When they are completed at $t_3$ and $t_4$, \texttt{FrontendNGINX} will trigger \texttt{PostStorageService} at $t_5$ to store the post content, which completes at $t_6$. Finally, \texttt{WriteHomeTimeline} is invoked at $t_7$ to start notifying the friends of the author and send the response to the client at $t_8$ acknowledging the compose post request has been done. We can identify three execution workflow patterns of components with the temporal information encoded in a trace:
\begin{itemize}[leftmargin=*,noitemsep,topsep=0pt]
	\item \emph{Parallel Execution}: The duration of two spans from the same parent overlaps significantly (e.g., \texttt{URLShortenService} and \texttt{MediaService}).
	\item \emph{Sequential Execution}: The duration of two spans from the same parent does not overlap (e.g., \texttt{URLShortenService} and \texttt{PostStorageService}).
	\item \emph{Background Execution}: The end time of the span exceeds the end time of its parent (e.g., \texttt{WriteHomeTimeline} and \texttt{FrontendNGINX}).
\end{itemize}

\textbf{Delay Injection.} Taking \texttt{URLShortenService} to offload as an example, the invocation from \texttt{FrontendNGINX} to it becomes long-distance with slower communication, as highlighted in Figure~\ref{fig:delay-injection}b (the orange bar). Let $\Delta$ be the delay, which will be approximated next. We can emulate the delay by updating the start time of \texttt{URLShortenService} from $t_1$ to $t'_1=t_1+\Delta$ and using its original execution duration (i.e., $t_3-t_1$ from Figure~\ref{fig:delay-injection}a) to update the end time to $t'_3=t'_1+(t_3-t_1)$. Delay injection does not end here. All downstream operations not running in parallel with \texttt{URLShortenService} should also be updated. The strategy is to recursively estimate how long it takes for the next operation to be triggered to update its start time and use the original execution duration to update the end time. Since \texttt{PostStorageService} runs sequentially after the two parallel operations which end at $\max{(t'_3, t_4)}$, we can approximate the time gap, and the corresponding span now starts at $t'_5=\max{(t'_3, t_4)}+[t_5-\max{(t_3, t_4)}]$ and ends at $t'_6=t'_5+(t_6-t_5)$ using its original execution duration (i.e., $t_6-t_5$). Similarly, we can update the start time of \texttt{WriteHomeTimeline} to $t'_7=t'_6+(t_7-t_6)$, and its end time is unimportant due to background execution with no impact on the API latency. Finally, the new response time is $t'_8=t'_7+(t_8-t_7)$, and the latency changes from $t_8-t_0$ to $t'_8-t_0$ after executing the migration plan. Note that \scheme{} delay injection is automated. No manual investigation has to be done. For each API, we obtain, e.g., $100$ traces from the telemetry server when its API latency stabilizes, repeat the delay injection on each trace to obtain the post-migration latency distribution (Figure~\ref{fig:footprint-dist0}), which matches the real one collected by actual migration, and extract the average latency. Let $\textsc{Lat}(\mathcal{A};\boldsymbol{p})$ be the new latency of API $\mathcal{A}$, given the migration plan $\boldsymbol{p}$, and $\textsc{Lat}(\mathcal{A})$ be the current one. The performance quality of a plan is the impact on the API latency, $\mathcal{Q}^{\text{Perf}}(\boldsymbol{p})=\frac{1}{\vert\boldsymbol{\mathcal{A}}\vert}\sum_{\mathcal{A}}\tau_\mathcal{A}\frac{\textsc{Lat}(\mathcal{A};\boldsymbol{p})}{\textsc{Lat}(\mathcal{A})}$, to be minimized, where $\vert\boldsymbol{\mathcal{A}}\vert$ is the total number of APIs, and $\tau_\mathcal{A}$ is the optional weight to reflect the importance of different APIs. Those critical APIs specified by the application owner are weighted $2\times$ more important by default.

\textbf{Network Footprint.} The cascade update originates from the delay $\Delta$. It depends on the data size to be transferred. \scheme{} must know, for a given API, how many bytes will be transmitted when one component triggers the other during the request and response. We call it the network footprint of an API. This seems to be available from Istio~\cite{istio} (see Section~\ref{sec:bg}), but it only captures the overall traffic between components aggregating requests of all APIs. We propose to learn the network footprint of each API by associating distributed traces and pairwise network metrics. The idea is to learn network footprints that can reconstruct real network traffic. From the telemetry server, we can obtain the total number of bytes $\boldsymbol{\mathcal{U}}^{\text{req}}_{c_i\rightarrow c_j}[t]$ being transferred from component $c_i$ to $c_j$ during requests at time $t$, which is a time window of, e.g., $5$ seconds. We also parse the traces generated by requests of API $\mathcal{A}$ collected within the time window $t$ to count the number of invocations from component $c_i$ to $c_j$, denoted as $\mathcal{I}^{\mathcal{A}}_{c_i\rightarrow c_j}[t]$. Then, we can jointly learn the request data size $d^{\mathcal{A}, \text{req}}_{c_i\rightarrow c_j}$ from component $c_i$ to $c_j$ for every user-facing API $\mathcal{A}$ by attempting to justify the actual traffic:
\begin{equation}
	\argmin_{d^{\mathcal{A}, \text{req}}_{c_i\rightarrow c_j} \;\forall\mathcal{A}}\sum_{t}\big[\boldsymbol{\mathcal{U}}^{ \text{req}}_{c_i\rightarrow c_j}[t]-\sum_{\mathcal{A}}\mathcal{I}^{\mathcal{A}}_{c_i\rightarrow c_j}[t]d^{\mathcal{A}, \text{req}}_{c_i\rightarrow c_j}\big]^2.
\end{equation}
The above optimization is run for each component pair, both requests and responses, to reconstruct the entire network footprint of each API (a visualization is given in Figure~\ref{fig:footprint-register}). While the number of required time windows to learn the footprints depends on the variation of data transfer size, the rule of thumb is to have at least ten for each parameter~\cite{harrell1984regression}. Then, the delay $\Delta$ of communications between component $c_i$ and $c_j$ due to the offloading of $c_j$ w.r.t. API $\mathcal{A}$ is
\begin{equation}
	\Delta=(\gamma_\text{after}-\gamma_\text{before})+(\nu_\text{after}-\nu_\text{before})(d^{\mathcal{A}, \text{req}}_{c_i\rightarrow c_j}+d^{\mathcal{A}, \text{resp}}_{c_i\rightarrow c_j}),
\end{equation}
where $\gamma_\text{after}$ and $\gamma_\text{before}$ are the average network latency between the locations of $c_i$ and $c_j$ after and before migration, respectively, while $\nu_\text{after}$ and $\nu_\text{before}$ are the bandwidth.

\textbf{Insights.} \scheme{} exploits the execution workflow to find better components for offloading, which cannot be done by considering the traffic between datacenters~\cite{sampaio2019improving,joseph2020intma,hu2019optimizing,han2020refining}. For instance, some components running in parallel may wait for another to complete (e.g., \texttt{MediaService} in Figure~\ref{fig:delay-injection}). Although the offloading of \texttt{URLShortenService} prolongs the communication (the orange bar), the impact on the end-to-end API latency is small (the pink bar). Also, the offloading of components running with a background workflow has no impact on the API latency at all. These can be captured by \scheme{}'s direct latency optimization.

\subsubsection{API Availability Modeling} 
The second aspect of quality is the disruption brought to the application during the migration process. A stateless component can be offloaded with minimal disruption by, e.g., using the rolling update mechanism, but stateful ones require data transfer to the new location. Thus, some services in the application may not be available until the migration is completed. Also, it can result in loss of cached data and lead to performance degradation after restart~\cite{deshpande2019caravel}. Hence, we aim to minimize such disruptions. Let $\textsc{SC}(\mathcal{A})$ be the set of stateful components used by API $\mathcal{A}$, which can be found by visiting the traces generated from API $\mathcal{A}$ during application learning. We quantify the disruption to API $\mathcal{A}$ given a plan $\boldsymbol{p}$:
\begin{equation}
	\mathcal{Q}^{\text{Avai}}(\mathcal{A};\boldsymbol{p})=
	\begin{cases}
		1, & \text{if } p_c\neq\ell_c\quad\forall c\in\textsc{SC}(\mathcal{A})\\
		0,              & \text{otherwise,}
	\end{cases}
\end{equation}
where $\ell_c$ is the original location of component $c$.
The availability quality of a plan is the weighted number of APIs under disruption, $\mathcal{Q}^{\text{Avai}}(\boldsymbol{p})=\sum_{\mathcal{A}}\tau_{\mathcal{A}}\mathcal{Q}^{\text{Avai}}(\mathcal{A};\boldsymbol{p})$, to be minimized.

\subsubsection{Cloud Hosting Cost Modeling}
The cloud hosting cost $\mathcal{Q}^{\text{Cost}}(\boldsymbol{p})$ of a plan $\boldsymbol{p}$ covers (i) compute, (ii) storage, and (iii) network traffic. Using a resource estimator (e.g., DeepRest~\cite{chow2022deeprest} used in this paper) and the migration plan, we can obtain the expected resource demands within a period of interest to be fulfilled by the cloud provider. Given the hardware specification of the node type from the cloud provider, we can calculate the number of nodes required over time (e.g., every ten minutes) to measure the compute-induced cost. For storage, since autoscaling is not a privilege of compute but is also supported in cloud storage, we incorporate such fine-grained pricing to calculate the amount of storage required from the cloud provider over time and the corresponding storage-induced cost. Finally, for network traffic, public clouds typically do not charge any data flow into their datacenters, but the egress traffic from the cloud can be expensive. We identify egress traffic according to the migration plan to approximate the traffic-induced cost. This pricing model is generalized to reflect the key characteristics of different public clouds~\cite{aws,azure,google}. We provide its details in Appendix~\ref{app:pricing}. The exact pricing can vary in real-time, and the application owner can provide the billing catalog, such as the query endpoint~\cite{gcpapi}, of the cloud provider of interest as a plugin for \scheme{} to adapt to the recommendations. Depending on the use case, the pricing model can be augmented to accommodate the traffic cost induced by restoring the placement for temporary migration. 

\subsection{Migration Plan Recommendation}
With the above modeling, \scheme{} aims to find the migration plan $\boldsymbol{p}$ with minimal performance impact on APIs $\mathcal{Q}^{\text{Perf}}$,  disruption to APIs $\mathcal{Q}^{\text{Avai}}$, and hosting cost $\mathcal{Q}^{\text{Cost}}$ while satisfying the constraints inferred from the migration preferences:
\begin{equation}\label{eq:obj1}
\hspace{-0.5em}	\begin{aligned}
		\min_{\boldsymbol{p}} \quad & \mathcal{Q}^{\text{Perf}}(\boldsymbol{p}) + \mathcal{Q}^{\text{Avai}}(\boldsymbol{p}) + \mathcal{Q}^{\text{Cost}}(\boldsymbol{p})\\
		\textrm{s.t.} \quad &p_c=\ell\quad\forall (c, \ell)\in\mathcal{M}_{\text{placement}}    \\
		& \max_t\;[\sum_{c\in\boldsymbol{\mathcal{C}}} (1-p_c)\tilde{\boldsymbol{\mathcal{U}}}^{r}_c[t]]\leq \mathcal{M}^{{r}}_\text{onprem-limit}\;\forall r\in{\boldsymbol{\mathcal{R}}}\\
		& \mathcal{Q}^{\text{Cost}}(\boldsymbol{p})\leq\mathcal{M}_{\text{budget}}\\
	\end{aligned}
\end{equation}
The first constraint allows the application owner to force a component $c$ to be at a fixed location $\ell$ for, e.g., regulatory compliance. Components not in $\boldsymbol{\mathcal{M}}_{\text{placement}}$ can be freely relocated. The second constraint allows setting the maximum usage $\mathcal{M}^{{r}}_\text{onprem-limit}$ for each on-prem resource $r$ of all resource types $\boldsymbol{\mathcal{R}}$, such as CPU, memory, and storage, and $\tilde{\boldsymbol{\mathcal{U}}}^{r}_c[t]$ denotes the expected usage of resource $r$ in component $c$ at time $t$. The default limits are extracted from cluster nodes but can be modified to reduce the allocated resources to the application. The last constraint ensures the cost does not exceed the budget $\mathcal{M}_{\text{budget}}$, which is infinity by default. 

One way to solve the above multi-objective optimization problem is to enumerate and calculate the quality of all possible plans, filter out those violating any constraint, and preserve the ones that excel. However, it is not scalable because of the gigantic search space. We can only visit a small number of candidates and hence demand an intelligent approach to explore only those worth our time.

\subsubsection{DRL-based Genetic Algorithm}
\scheme{} takes an evolutionary approach~\cite{van1998multiobjective} with genetic algorithms (GAs)~\cite{mirjalili2019genetic} to find high-quality plans by visiting a small number of carefully selected candidates. As depicted in Figure~\ref{fig:overview}, \textcircled{1} it starts with randomly choosing a small population (e.g., $100$) of plans. \textcircled{2} Each plan goes through the quality assessment process by calculating its incurred API performance $\mathcal{Q}^{\text{Perf}}$, API availability $\mathcal{Q}^{\text{Avai}}$, and hosting cost $\mathcal{Q}^{\text{Cost}}$.  \textcircled{3} Then, we only keep the subset of ``Pareto optimal''~\cite{horn1994niched} plans, meaning that for each plan in the subset, one could not find another plan better in one aspect of quality (e.g., cost) without sacrificing the other (e.g., performance). This is mandatory because the three quality indicators are contradictory. We expect to produce a set of migration plans with different trade-offs. \textcircled{4} Next, we select pairs of plans with diverse strengths (e.g., one with a cheap hosting cost and one with a low-performance impact). \textcircled{5} For each pair, we combine them (crossover) and produce the so-called offspring plan to be added to the population. The above repeats from \textcircled{2} for several generations, and the crux is to generate high-quality offspring that can outperform their parents. They will replace weaker ones in the population in the upcoming generation (i.e., in \textcircled{3}). As the generation progresses, the population of plans gets more and more competitive, and the plans that ``survive'' this rigorous selection in the last generation are recommended migration plans optimizing Equation~\ref{eq:obj1}. They are with different trade-offs for selection based on the business need. 

Existing approaches create offspring by randomly combining the parents~\cite{guerrero2018resource,ding2022kubernetes}. We argue that a more goal-driven crossover can boost the likelihood of producing a child surpassing both of its parents, and as a result, the evolution of the population can be accelerated. We introduce an optimization approach to conduct intelligent crossover, formulating it as a learning problem with dual goals: (i) the produced offspring plan should be a feasible plan satisfying all constraints, and (ii) the produced offspring plan should outperform its parents in as many quality aspects as possible. The first goal is particularly important because one should not waste time on any plan that cannot be deployed (e.g., on-prem components exceed the resource limits).

\scheme{} uses non-dominated sorting, crowding distance, and binary tournament in NSGA-II~\cite{deb2002fast} to select pairs of parent plans for crossover. We refer the readers to~\cite{deb2002fast} for details. Given two parent plans $(\boldsymbol{p}^i, \boldsymbol{p}^j)$ from the tournament selection, we desire an intelligent agent $\Lambda_{\boldsymbol{\theta}}$ that takes the concatenated vector of both parents as input and returns the child plan $\Lambda_{\boldsymbol{\theta}}(\boldsymbol{p}^i\vert\vert\boldsymbol{p}^j)$ that is better than both parents in ideally all quality aspects. This can be done by training a neural network, parameterized by $\boldsymbol{\theta}$, from a dataset $\boldsymbol{\mathcal{D}}$ of a small number of plans such that with iterative learning, it learns from the cases where its crossover leads to a high-quality child and avoids making the same mistake when it does not. However, based on Section~\ref{sec:modeling}, the quality indicators are non-differentiable, and hence $\Lambda_{\boldsymbol{\theta}}$ cannot be directly optimized by backward propagation~\cite{lecun2015deep}. Inspired by the recent advances in deep reinforcement learning (DRL) to train intelligent agents with a non-differentiable objective~\cite{arulkumaran2017deep,xie2018environment,hui2021learning}, we reformulate the learning as a reward-driven problem to train $\Lambda_{\boldsymbol{\theta}}$ to predict a probability distribution of high-quality child plans given two parents $(\boldsymbol{p}^i, \boldsymbol{p}^j)$ with $\arg\max_{\boldsymbol{\theta}}\mathop{\mathbbm{E}}_{(\tilde{\boldsymbol{p}}^i, \tilde{\boldsymbol{p}}^j)\sim\boldsymbol{\mathcal{D}}}\textsc{Reward}(\boldsymbol{p};\boldsymbol{p}^i,\boldsymbol{p}^j)$, where
\begin{equation}\label{eq:reward}
	\begin{split}
		&\textsc{Reward}(\boldsymbol{p};\boldsymbol{p}^i,\boldsymbol{p}^j)\\
\hspace{-0.8em}		=&(-1)^{1-\lambda(\boldsymbol{p})}\sum_{\mathcal{Q}\in\{\mathcal{Q}^{\text{Perf}}, \mathcal{Q}^{\text{Avai}}, \mathcal{Q}^{\text{Cost}}\}}\mathbbm{I}[\min_{k\in\{i,j\}}\mathcal{Q}(\boldsymbol{p}^k)>\mathcal{Q}(\boldsymbol{p})],
	\end{split}
\end{equation}
with $\lambda(\boldsymbol{p})$ checking the feasibility (i.e., satisfying all constraints in Equation~\ref{eq:obj1}) of the child plan $\boldsymbol{p}\sim\Lambda_{\boldsymbol{\theta}}(\boldsymbol{p}^i\vert\vert\boldsymbol{p}^j)$, which is $1$ if feasible and $0$ otherwise, and $\mathbbm{I}[\cdot]$ being a binary function returning $1$ if the condition is true.

The reward function guides the learning of $\Lambda_{\boldsymbol{\theta}}$, which needs to make crossover decisions to maximize the reward. It requires providing appropriate learning signals, especially penalizing unwanted decisions. We make three design considerations. First, to produce a child that can maximize the reward, it must be a feasible solution because the term, $(-1)^{1-\lambda(\boldsymbol{p})}$, in Equation~\ref{eq:reward} negates the reward if the plan does not satisfy all constraints. Second, the reward function encourages the crossover process to produce a child with better quality than its parents in as many aspects as possible. Note that \scheme{} does not directly maximize the margin of improvement in different aspects. This is because their differences in scale can mislead the learning to treat one aspect to dominate (e.g., availability). We require $\Lambda_{\boldsymbol{\theta}}$ to discover high-quality plans equally in all aspects. Third, with the concatenated vector of both parents as input (i.e., the state space in DRL), we model $\Lambda_{\boldsymbol{\theta}}$ to output a probability distribution (i.e., the action space in DRL). The sampling process has the same spirit as the mutation process in GAs~\cite{mirjalili2019genetic} to introduce diversity to the population. With the reward function, we use the actor-critic algorithm in DRL~\cite{haarnoja2018soft}, popularly used in AI applications~\cite{su2017sample,han2020actor,leng2021actor}, to train $\Lambda_{\boldsymbol{\theta}}$. At convergence, $\Lambda_{\boldsymbol{\theta}}$ can conduct crossover on two given plans to generate a child that is feasible and better than both parents.

\subsubsection{Hierarchical Post-processing}
\begin{figure}
	\centering
	\begin{subfigure}[b]{0.49\linewidth}
		\centering
		\includegraphics[width=0.92\linewidth]{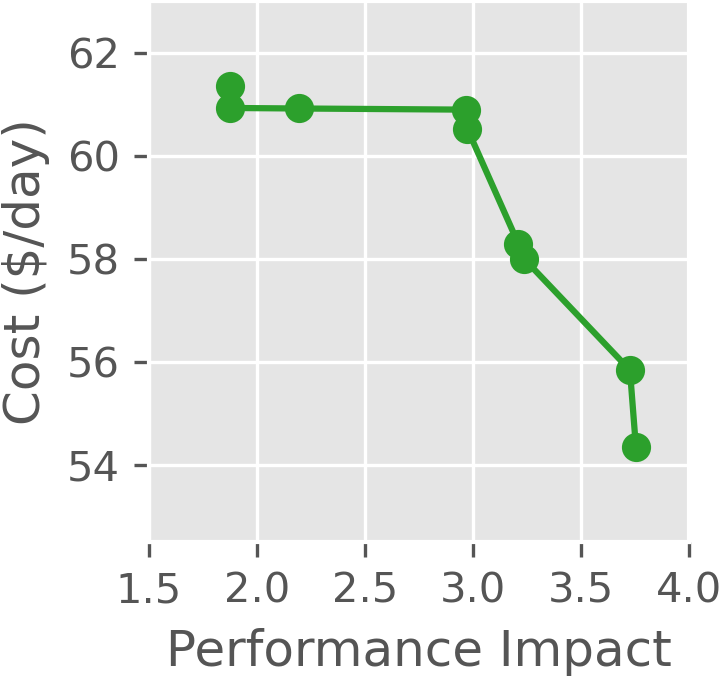}
		\caption{Pareto Front}\label{fig:atlas-front}
	\end{subfigure}
	\hfill
	\begin{subfigure}[b]{0.49\linewidth}
		\centering
		\includegraphics[width=0.92\linewidth]{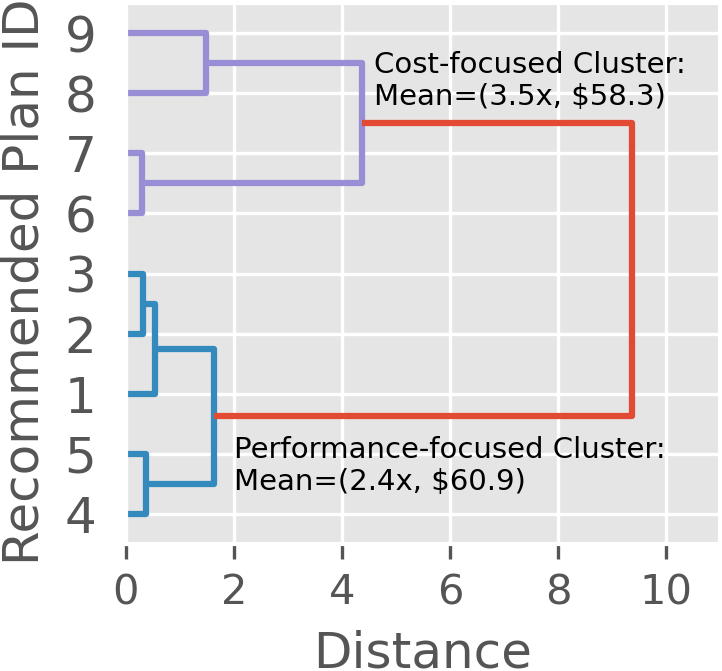}
		\caption{Hierarchical Representation}\label{fig:atlas-dendrogram}
	\end{subfigure}
	\caption{The nine Pareto optimal plans recommended by \scheme{} with a hierarchical representation to aid the plan selection.}
\end{figure}
Our DRL-based genetic algorithm outputs a list of Pareto optimal plans, often presented as a Pareto front. Figure~\ref{fig:atlas-front} gives an example with only two objectives: performance and cost. Each point corresponds to a recommended plan. In practice, such a representation, especially in three or more dimensions, is not friendly for the application owner to select the option that fits their needs. Hence, we use hierarchical clustering~\cite{johnson1967hierarchical} to organize the recommended plans with a tree structure named dendrogram~\cite{han2011data}, where the application owner is first given a few plans representing clusters with different high-level characteristics (e.g., performance-focused or cost-focused), as shown in Figure~\ref{fig:atlas-dendrogram}. Then, more fine-grained clusters are presented for selection until reaching the leaves, representing the actual plans. This helps the application owner strategically narrow down the options and avoid being overwhelmed by the number of possibilities at the beginning.

\subsection{Post-migration Monitoring}\label{sec:monitoring}
The post-migration monitoring stage aims at proactively initiating a new round of recommendations when better plans can be offered. This can happen due to both external and internal factors. External factors are related to user behavior changes. For instance, the API composition received by the application can change over time~\cite{chow2022deeprest}, making some components more popular than others. Internal factors refer to the changes in API footprints. 
\begin{figure}
	\centering
	\includegraphics[width=0.95\linewidth]{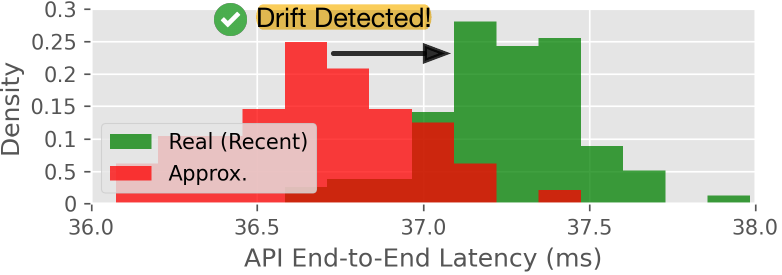}
	\caption{\scheme{} can detect outdated footprints and proactively recommend better placement plans for migration.}\label{fig:footprint-dist}
\end{figure}
One can expect that when the application gets popular with, e.g., more users sharing posts, the response to the \texttt{/homeTimeline} request contains more posts and is larger in size. This drift can lead to inaccuracy in the API performance estimation, as shown in Figure~\ref{fig:footprint-dist}. Since the executed plan was selected based on the API performance that is no longer valid, it is crucial to detect such an event and recommend new plans proactively.

To detect such changes while accommodating normal uncertainties, we use a statistical approach. Let $b^{\mathcal{A}}_{\text{approx}}$ be the latency distribution of API $\mathcal{A}$ approximated by \scheme{} in the last round of recommendations. We can collect recent traces and obtain the most up-to-date distribution $b^{\mathcal{A}}_{\text{real-recent}}$ to verify the validity of $b^{\mathcal{A}}_{\text{approx}}$ using Kullback-Leibler divergence ($D_{\text{KL}}$)~\cite{Joyce2011}, where statistically similar distributions have a lower score. It has no assumption about the family of distributions and can be used to detect any distributional changes. However, it is difficult to determine whether the change is significant without a reference, especially for KL divergence with no upper bound (range from zero to infinity). Hence, we take the latency distribution of the same API $\mathcal{A}$ captured in the previous round $b^{\mathcal{A}}_{\text{real}}$ as a baseline to determine the significance of distribution changes. For instance, the latency of \texttt{/homeTimeline} in Figure~\ref{fig:footprint-dist0} gives a baseline of $D_{\text{KL}}(b^{\mathcal{A}}_{\text{real}}, b^{\mathcal{A}}_{\text{approx}})=0.47$, which becomes $D_{\text{KL}}(b^{\mathcal{A}}_{\text{real}}, b^{\mathcal{A}}_{\text{real-recent}})=6.09$ in Figure~\ref{fig:footprint-dist} when its API performance estimation is no longer valid due to the footprint changes. In information theory, the recent latency distribution $b^{\mathcal{A}}_{\text{real-recent}}$ loses $13\times$ information in approximating $b^{\mathcal{A}}_{\text{real}}$. A new round can begin to learn updated network footprints, conduct delay injection, and run DRL-based genetic algorithm for recommendations.

\section{Experimental Evaluation}
\subsection{Experiment Setup}
\begin{figure}
	\centering
	\includegraphics[width=\linewidth]{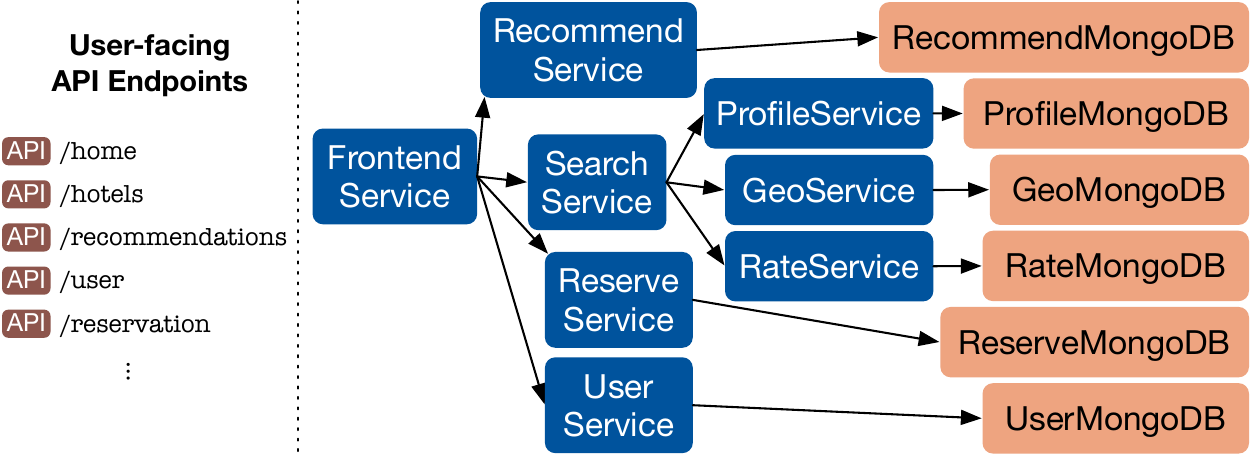}
	\caption{The hotel reservation system from DeathStarBench~\cite{gan2019open}.}
	\label{hr-arch}
\end{figure}

\textbf{Microservice Applications.} We evaluate \scheme{} on two applications from DeathStarBench~\cite{gan2019open}: a social network and a hotel reservation system. The social network has $23$ stateless and $6$ stateful components, offering $9$ user-facing APIs (Figure~\ref{sn-arch}). 
The hotel reservation system has $12$ stateless and $6$ stateful components with $5$ user-facing APIs (Figure~\ref{hr-arch}). They cover diverse workflow patterns. By default, we use the social network for evaluation due to its higher complexity.

\textbf{Workload Generation.} We generate workloads with real-world behaviors. The network graph and post contents are from real-world Facebook datasets with realistic user interactions~\cite{rossi2015network}. The media is from the INRIA dataset~\cite{dalal2005histograms}, having pictures of people with various resolutions and sizes. For the hotel reservation system, we follow the same setting as in~\cite{gan2019open}. Our Locust-based~\cite{locust} generator simulates one-day traffic in five minutes, where each day has two peak hours (e.g., lunchtime and late evening) to resemble real-world phenomena. API requests are sent according to real-world distributions with variations from day to day to follow non-deterministic properties in practice~\cite{kwak2010twitter}.

\textbf{System Setup and Hyperparameters.} All microservices are deployed in separate Docker containers orchestrated by Kubernetes~\cite{k8s}. We install the most commonly-used telemetry tools, including Jaeger~\cite{jaeger} for distributed tracing, cAdvisor~\cite{cadvisor} for component-focused resource monitoring, and Istio~\cite{istio} for pairwise network monitoring, with their default configuration. We use the same hyperparameter setting for \scheme{} on both applications with two days of data for application learning. The actor network in the DRL-based genetic algorithm has three ReLU layers with $128$ hidden units and is trained for $1,000$ iterations with the Adam optimizer~\cite{kingma2014adam}. 

\textbf{Hybrid Cloud Setup.} We conduct systematic experiments by initially placing all containers on a ten-node (on-prem) cluster located in Wisconsin provided by CloudLab~\cite{Duplyakin+:ATC19}, where each node has two Intel E5-2660v3 10-core CPUs at 2.60GHz, 160GB memory, 480GB SSD, and Dual-port Intel X520-DA2 10Gb NIC. Components are offloaded to a datacenter located in Massachusetts (public cloud), where each node has two Intel E5-2660v3 10-core CPUs at 2.60GHz or more, 256GB memory, 900GB SSD, and Dual-port Solarflare SFC9120 10Gb Ethernet NIC. The average latency and bandwidth between collocated nodes are 0.168ms and 941Mbps, respectively. For inter-datacenter communications, the average latency and bandwidth are 23.015ms and 921Mbps, respectively. We consider the scenario where the application has to serve API traffic with $5\times$ more users than ever and consumes resources beyond the capacity of the on-prem infrastructure, resembling Figure~\ref{fig:intro-overload}. The peak CPU utilization in our experiment reaches $264\%$. We assume user-generated data in \texttt{UserMongoDB}, \texttt{PostStorageMongoDB}, and \texttt{MediaMongoDB} of the social network and \texttt{UserMongoDB} and \texttt{ReserveMongoDB} of the hotel reservation system cannot be relocated to mimic real-world scenarios for regulatory compliance. All other components can be offloaded if recommended. We use DeepRest~\cite{chow2022deeprest} to estimate the expected resources to serve the traffic and \scheme{} to recommend migration plans.

\subsection{Recommendation Quality Analysis}\label{sec:exp-quality}
To highlight the advantages of \scheme{}, we compare it with six approaches in two categories. The first category produces one recommended migration plan for the application owner to execute. We implement two state-of-the-art approaches: REMaP~\cite{sampaio2019improving} and IntMA~\cite{joseph2020intma}. Similar to \scheme{}, they do not require any stress test on a separate cluster for a fair comparison. Both approaches minimize interactions between datacenters, where IntMA considers the overall traffic size between component pairs, and REMaP reduces the traffic size as well as the number of message exchanges. Furthermore, we include two greedy baselines to offload the busiest (largest) or the least busy (smallest) components until the on-prem cluster can fulfill the remaining ones~\cite{guo2012seagull}. Since our workloads are CPU-intensive, we use CPU usage to reflect the busyness. The second category recommends a set of migration plans with different trade-offs. We implement an affinity-based approach that uses NSGA-II~\cite{deb2002fast} to find migration plans with two optimization objectives: (i) minimizing the traffic size between datacenters (implying performance) and (ii) minimizing cloud hosting costs. Here, we use the same cost model as \scheme{}. This approach is representative of existing methods~\cite{hu2019optimizing,han2020refining,aznavouridis2022micro,ding2022kubernetes,guerrero2018resource}. We further include a baseline for this category using a random search. For a fair comparison, both approaches in this category, including \scheme{}, only visit $10,000$ plans (i.e., $0.0019\%$ of all possibilities), and both \scheme{} and the affinity-based GA have a population of $100$.  We only consider Pareto optimal plans~\cite{horn1994niched}: $23$ from \scheme{}, $6$ from affinity-based GA, and $10$ from random search. 

\subsubsection{Comparison with Single-plan Approaches}\label{sec:exp-single-plan}
\begin{figure}
	\centering
	\includegraphics[width=\linewidth]{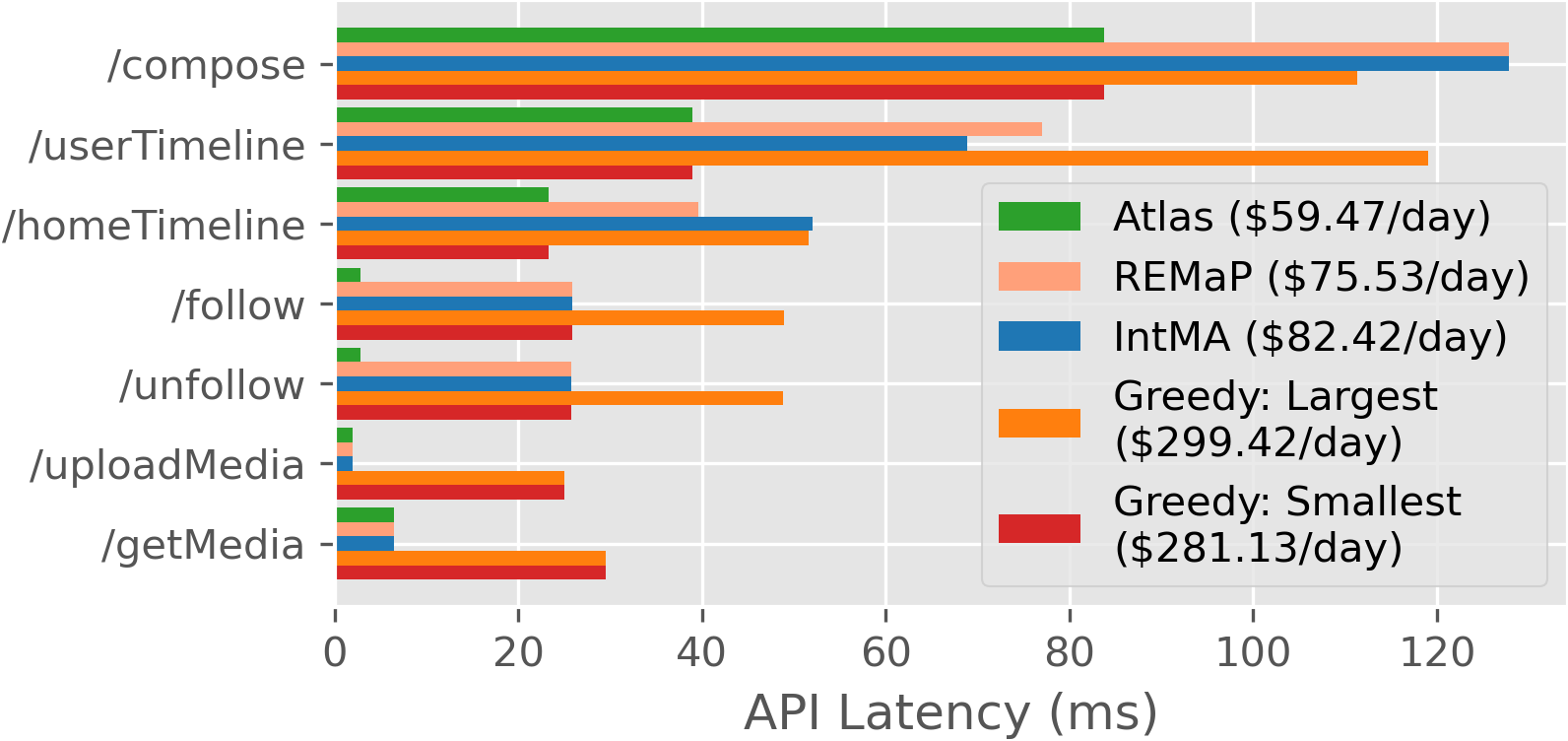}
	\caption{The state-of-the-art single-plan approaches (REMaP~\cite{sampaio2019improving} and IntMA~\cite{joseph2020intma}) and greedy approaches are more costly and lead to much worse API latency than \scheme{}.}
	\label{fig:greedy}
\end{figure}

We compare the performance-optimized option from \scheme{} with four single-plan approaches in Figure~\ref{fig:greedy} in terms of the latency of seven APIs and the cost per day. 
\begin{figure*}
\centering
\begin{subfigure}[b]{0.33\textwidth}
	\centering
	\includegraphics[width=\linewidth]{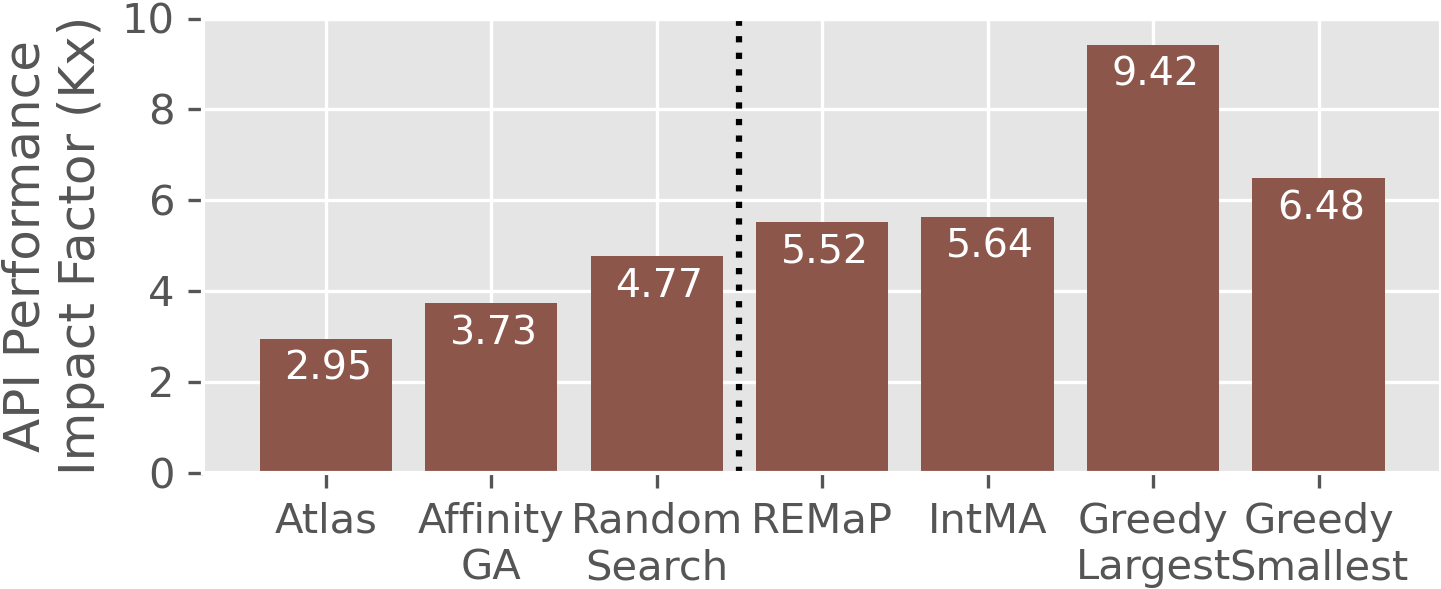}
	\caption{\bf \underline{API Performance Impact}}\label{fig:marginal-perf-a}
\end{subfigure}
\hfill
\begin{subfigure}[b]{0.33\textwidth}
	\centering
	\includegraphics[width=\linewidth]{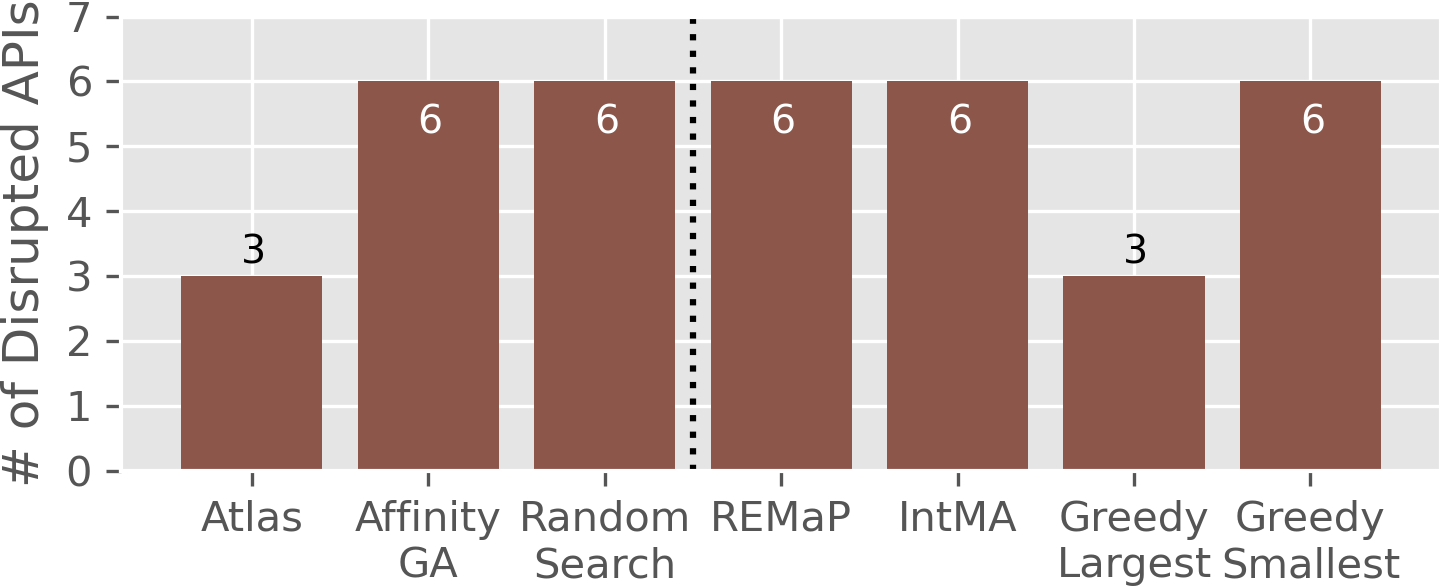}
	\caption{API Disruption (Availability)}
\end{subfigure}
\hfill
\begin{subfigure}[b]{0.33\textwidth}
	\centering
	\includegraphics[width=\linewidth]{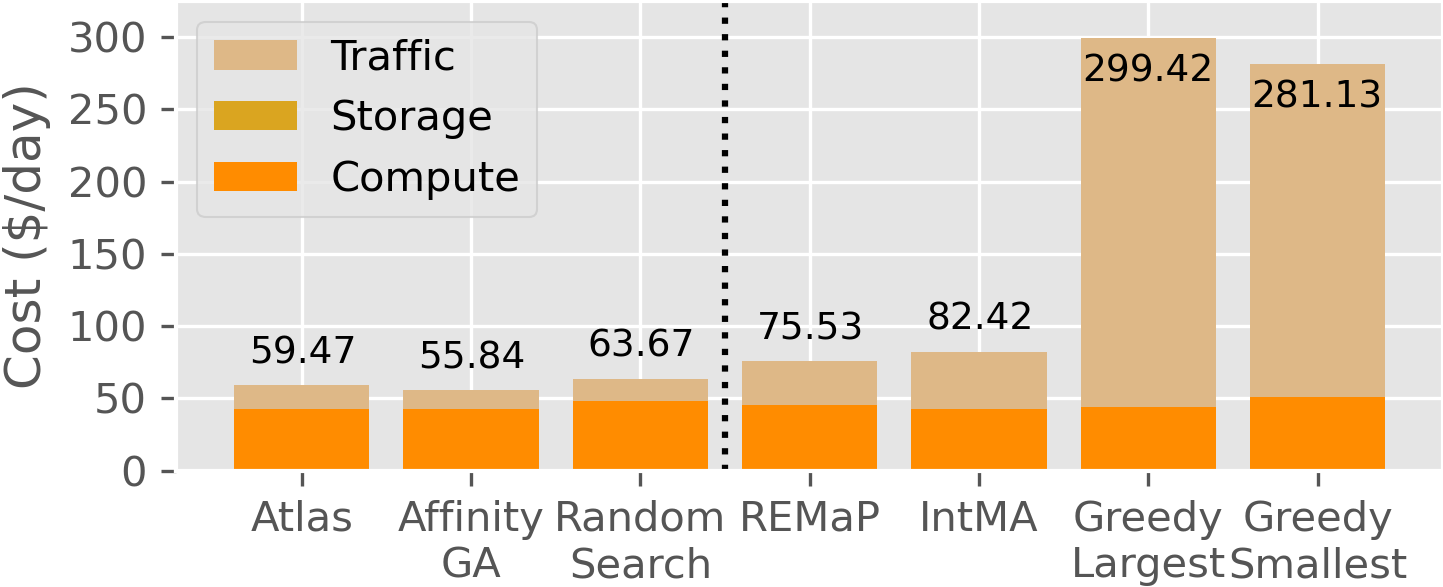}
	\caption{Cost}\label{fig:marginal-perf-b}
\end{subfigure}\vspace{-0.5em}
\caption{
	Given the performance-optimized plans from \scheme{} and six baselines, \scheme{} can find the plan with the least impact on API performance (a). While not the primary goal, the plan also incurs less disruption to APIs (b) and remains cheap (c).
}\label{fig:marginal-perf}
\end{figure*}
\begin{figure*}
\centering
\begin{subfigure}[b]{0.33\textwidth}
	\centering
	\includegraphics[width=\linewidth]{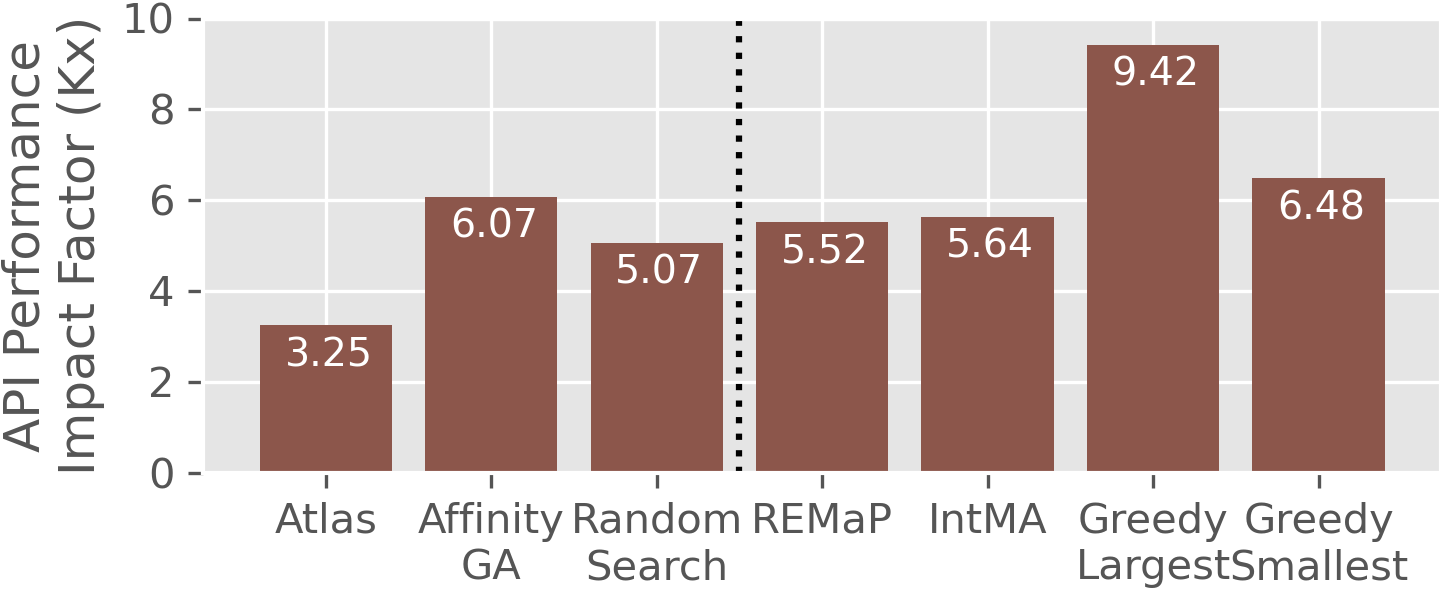}
	\caption{API Performance Impact}
\end{subfigure}
\hfill
\begin{subfigure}[b]{0.33\textwidth}
	\centering
	\includegraphics[width=\linewidth]{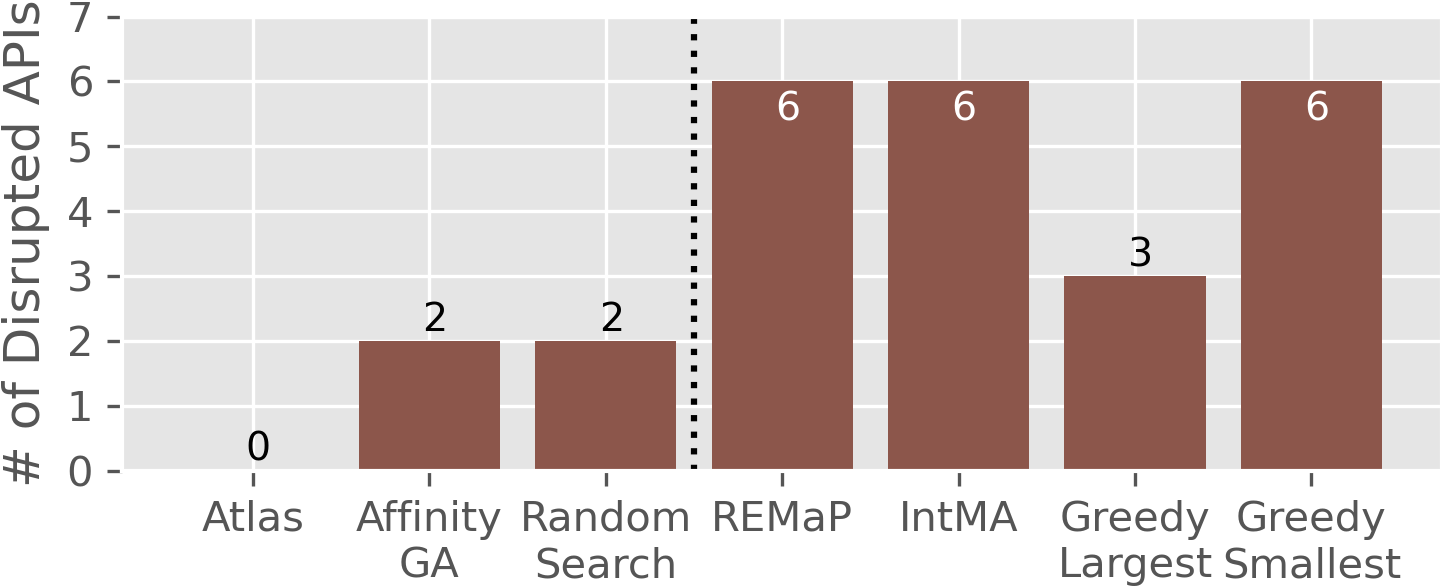}
	\caption{\bf \underline{API Disruption (Availability)}}
\end{subfigure}
\hfill
\begin{subfigure}[b]{0.33\textwidth}
	\centering
	\includegraphics[width=\linewidth]{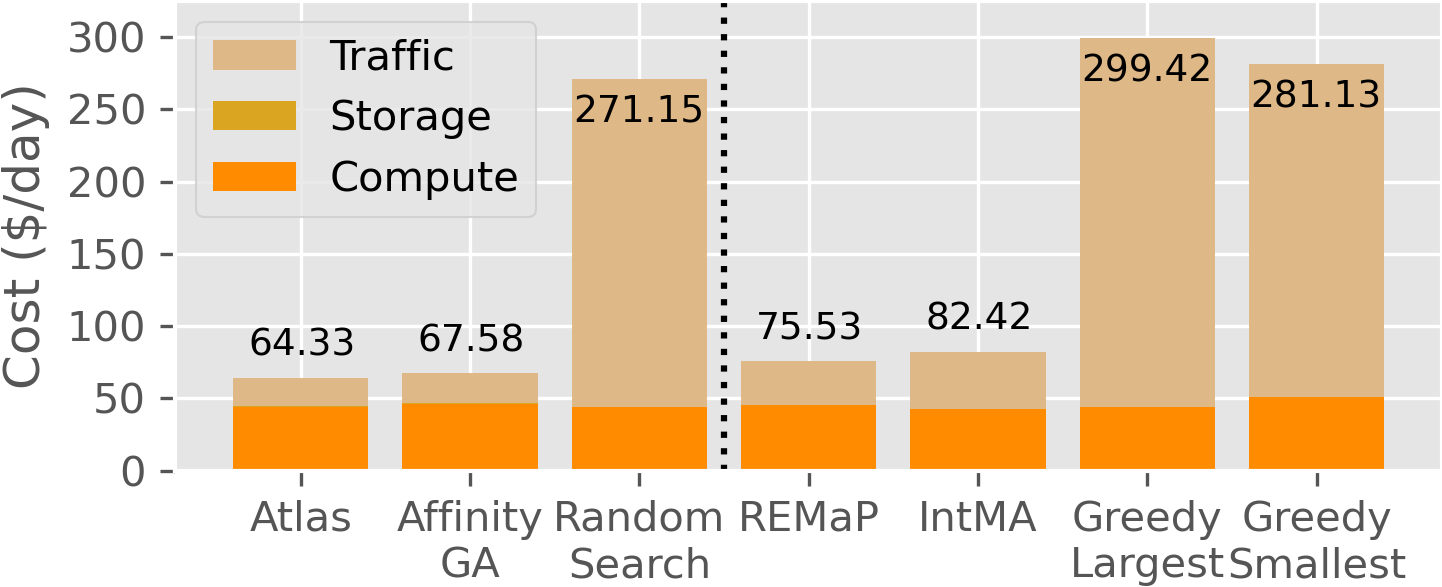}
	\caption{Cost}\label{fig:marginal-availability-b}
\end{subfigure}\vspace{-0.5em}
\caption{
	The optimization of API disruption (availability) in \scheme{} can find the plan with the least number of APIs with service disruption (b). The plan also incurs less API performance impact (a) and is cheaper (c) than the other six approaches.
}\label{fig:marginal-availability}
\end{figure*}
\begin{figure*}
\centering
\begin{subfigure}[b]{0.33\textwidth}
	\centering
	\includegraphics[width=\linewidth]{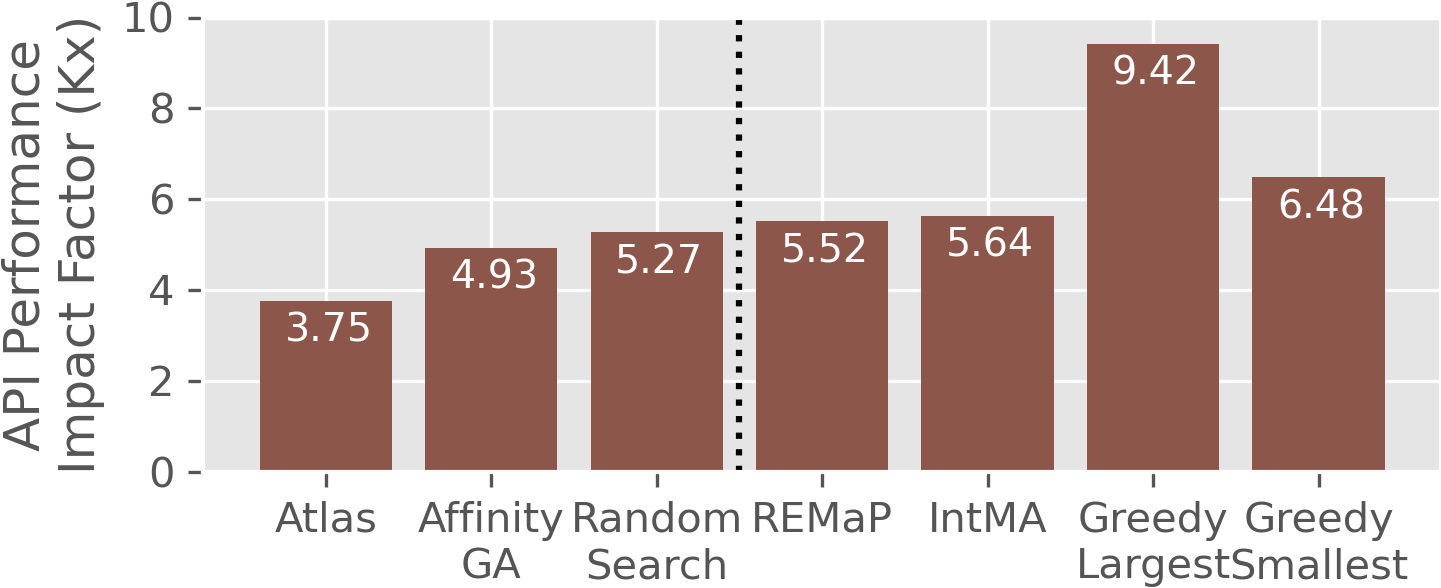}
	\caption{API Performance Impact}
\end{subfigure}
\hfill
\begin{subfigure}[b]{0.33\textwidth}
	\centering
	\includegraphics[width=\linewidth]{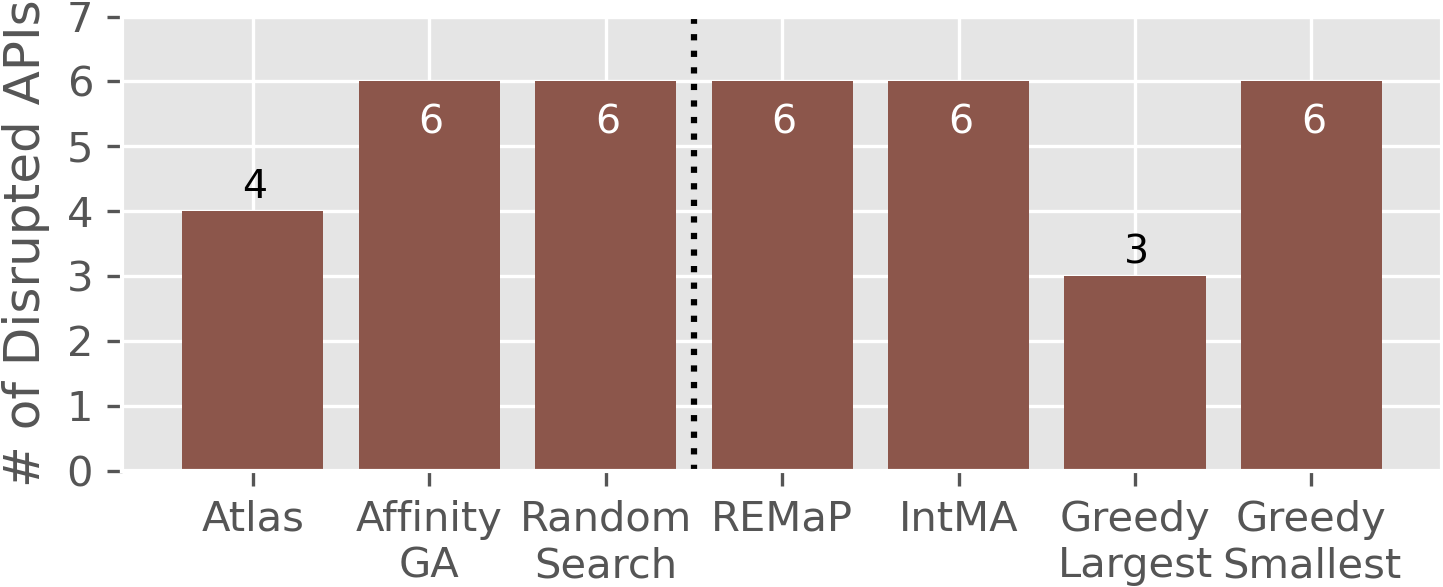}
	\caption{API Disruption (Availability)}
\end{subfigure}
\hfill
\begin{subfigure}[b]{0.33\textwidth}
	\centering
	\includegraphics[width=\linewidth]{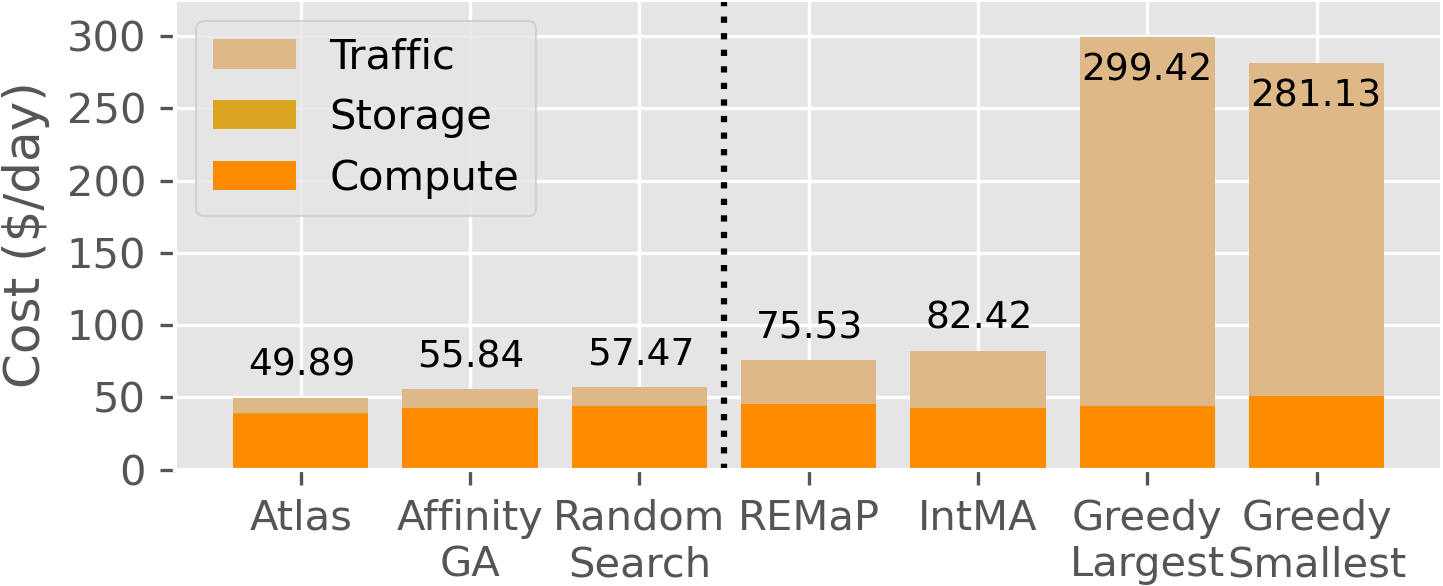}
	\caption{\bf \underline{Cost}}
\end{subfigure}\vspace{-0.5em}
\caption{
	Considering the cost-optimized plans, \scheme{} is the cheapest compared with the six approaches (c). The cheapest plan from \scheme{} also has less API performance impact (a) and limited disruption to APIs (b).
}\label{fig:marginal-cost}
\end{figure*}
The latency of each API under \scheme{}'s migration plan is consistently the lowest. The two greedy methods focus on the resource consumption of \emph{individual} components. They can create a feasible plan, but the API latency is significantly worsened by $2.69\times$ to $13.88\times$ (compared with $1.00\times$ to $4.74\times$ by \scheme{}) because there exists at least one non-background inter-datacenter communication to serve a request after the migration. Offloading the smallest components tends to better preserve API latency as they may be rarely triggered. \scheme{} outperforms REMaP and IntMA because it considers the workflow patterns of APIs. For instance, offloading multiple components being triggered in parallel may not lead to minimal interactions between datacenters but can create room to fit other components without additional latency. 

We also observe that even though the cost is not the primary objective for this plan, \scheme{} is also the cheapest ($\$59.47$ per day). The component-focused greedy approaches ignore the interaction between components. They assign components with intense data transfer in separate clusters, incurring significant egress traffic costs, and are  $5\times$ more expensive than \scheme{}. For REMaP and IntMA, while their objective is to reduce traffic (part of the cost) between datacenters, the search for the plan to minimize the objective is based on a simple heuristic, which leads to suboptimal solutions ($1.27\times$ for REMaP and $1.38\times$ for IntMA). \scheme{} outperforms single-plan approaches in terms of both API performance and cost. It shows the importance of recommending migration plans considering multiple aspects of application-level quality but not individual components or their pairwise interactions.

\subsubsection{Comparison with Multi-plan Approaches} 
The quality-driven approaches often lead to multiple recommendations with different trade-offs due to contradictory objectives (e.g., lower cost means higher performance impact).
We first analyze the recommended plan that excels at each aspect of quality: the performance-optimized plan in Figure~\ref{fig:marginal-perf}, the availability-optimized plan in Figure~\ref{fig:marginal-availability}, and the cost-optimized plan in Figure~\ref{fig:marginal-cost}. For example, the performance-optimized plans evaluated in Figure~\ref{fig:marginal-perf} are the best performance plan from each method. It can be selected by an application owner with an unlimited budget and does not mind having temporary service disruption during the migration process but strives to provide the best user experience (i.e., responsiveness). We compare \scheme{} with two multi-plan approaches and also include the four single-plan methods even though our evaluation in Section~\ref{sec:exp-single-plan} has already exposed their weaknesses, and they only recommend one plan.

Focusing on the performance-optimized option (Figure~\ref{fig:marginal-perf}), we define the API performance impact factor to be an evaluation metric for performance such that a factor of $K$ means APIs on a hybrid cloud setting based on a migration plan are, on average, $K$ times slower than before (i.e., when all components are on-prem and without resource stress). Intuitively, our goal is to minimize $K$.  \scheme{} is the most effective in choosing components to offload. Under \scheme{}'s recommendation, an API is only $2.95\times$ slower on average, much better than the second-best option by affinity-based GA, leading to $3.73\times$ slower API responsiveness (i.e., $20.91\%$ improvement by \scheme{}), which has no API awareness and cannot be customized to favor critical APIs (see Section~\ref{sec:exp-person}). Since cloud nodes often come with a fixed resource granularity (e.g., four CPU cores), we observe that \scheme{}'s tactic is to take advantage of the unused resources and offload more components, especially those with a background or parallel execution workflow that does not lead to a significant impact on end-to-end latency. They may slightly increase the traffic-induced cost (Figure~\ref{fig:marginal-perf-b}) but can improve the API performance, which is the primary objective for this performance-optimized case. The improvement would be the most prominent when the application has APIs with complex workflows, allowing \scheme{} to offload components that can lead to minimal performance impact and create room for those which cannot and should remain on-prem. Even though the plan is selected w.r.t. performance, \scheme{}'s recommendation is also high-quality in terms of availability and cost. This observation can also be made in other cases (Figure~\ref{fig:marginal-availability} and Figure~\ref{fig:marginal-cost}). For the random search, the quality is purely by chance, and it is unlikely to sample a high-quality one from over $500$ million possibilities. 
\begin{figure}
	\centering
	\includegraphics[width=0.86\linewidth]{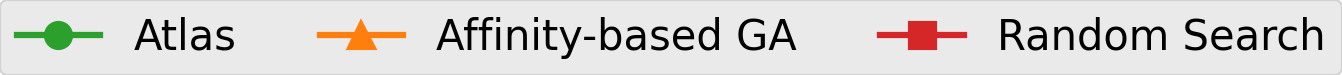}\vspace{0.5em}
	\begin{subfigure}[b]{0.49\linewidth}
		\centering
		\includegraphics[width=0.95\linewidth]{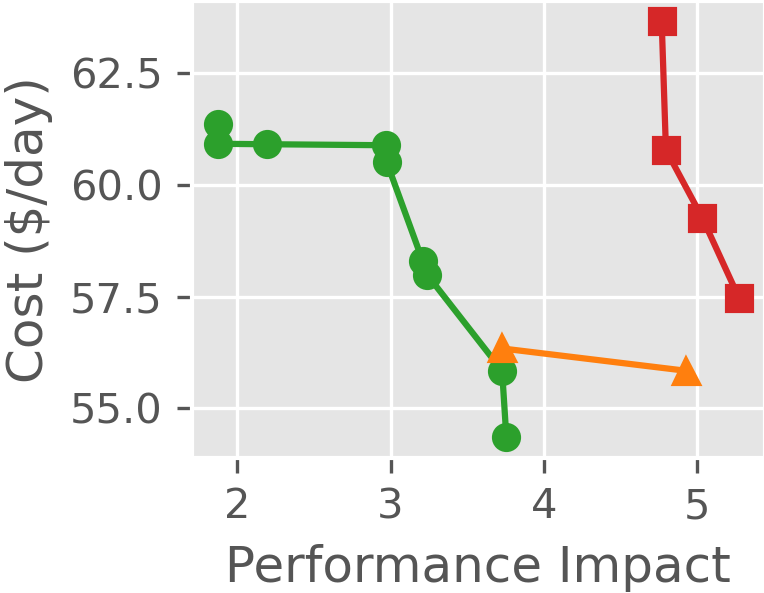}
		\caption{Social Network}\label{fig:exp-fronts-a}
	\end{subfigure}
	\hfill
	\begin{subfigure}[b]{0.49\linewidth}
		\centering
		\includegraphics[width=0.95\linewidth]{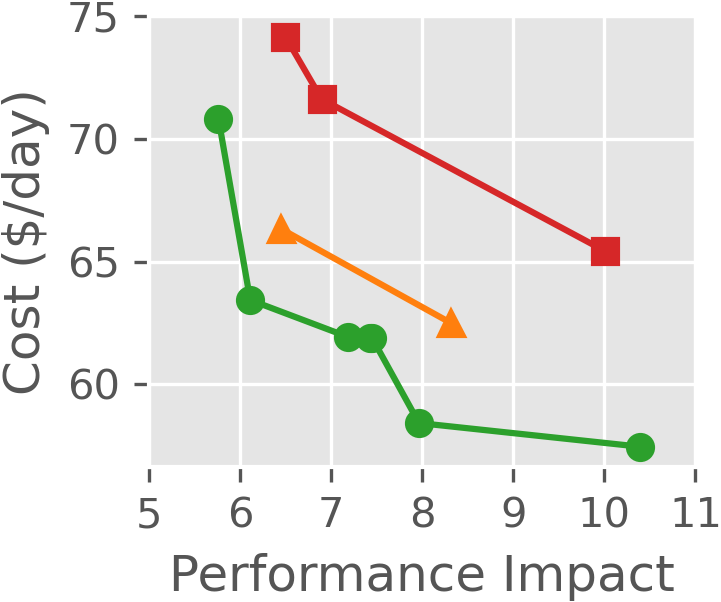}
		\caption{Hotel Reservation}\label{fig:exp-fronts-b}
	\end{subfigure}
	\caption{\scheme{} offers the best trade-off between different objectives. For any plan from other approaches, it can find a better one.}\label{fig:exp-fronts}
\end{figure}

The availability optimization is unique in \scheme{}. As shown in Figure~\ref{fig:marginal-availability}, \scheme{} is the only solution that can lead to the least disruption to APIs, where affinity-based GA and random search can only achieve it by chance (including all four single-plan methods). Recall that these two approaches recommend six and ten plans, respectively. The best plan from their recommendations still leads to two APIs suffering from disruption, but the application owner has the choice to avoid it by consulting \scheme{}. Finally, for the cost-optimized plan (Figure~\ref{fig:marginal-cost}), the storage-induced cost is negligible. An interesting observation can be made by comparing the cost of \scheme{} and affinity-based GA, where we use the same cost model. They both avoid the expensive egress traffic. A poor choice by greedy approaches can increase the daily cost to over $\$281.13$. \scheme{} costs only $\$49.89$, less than $\$55.84$ by affinity-based GA (the second cheapest among all six comparison schemes), leading to a $10.66\%$ improvement. This can be attributed to \scheme{}'s DRL approach. We defer the detailed analysis of it to Section~\ref{sec:exp-ablation}.

Application owners can choose a plan based on their business needs with different preferences between performance, availability, and cost. The role of \scheme{} is to offer the best plans available for selection. Figure~\ref{fig:exp-fronts} shows the Pareto fronts comparing three approaches for two applications. Recall that each point along the front corresponds to a recommended plan. For visualization simplicity, we optimize only two quality indicators and show the two-dimensional results on cost (y-axis) and performance impact (x-axis). For both applications, given \emph{any} plan recommended by either affinity-based GA or random search, \scheme{} can find another plan better in at least one aspect without sacrificing the others. Hence, \scheme{} provides the best trade-offs for the application owner to examine and choose the one that fits their needs. 

\subsection{API-aware Personalized Recommendations}\label{sec:exp-person}
\begin{figure}
	\centering
	\includegraphics[width=\linewidth]{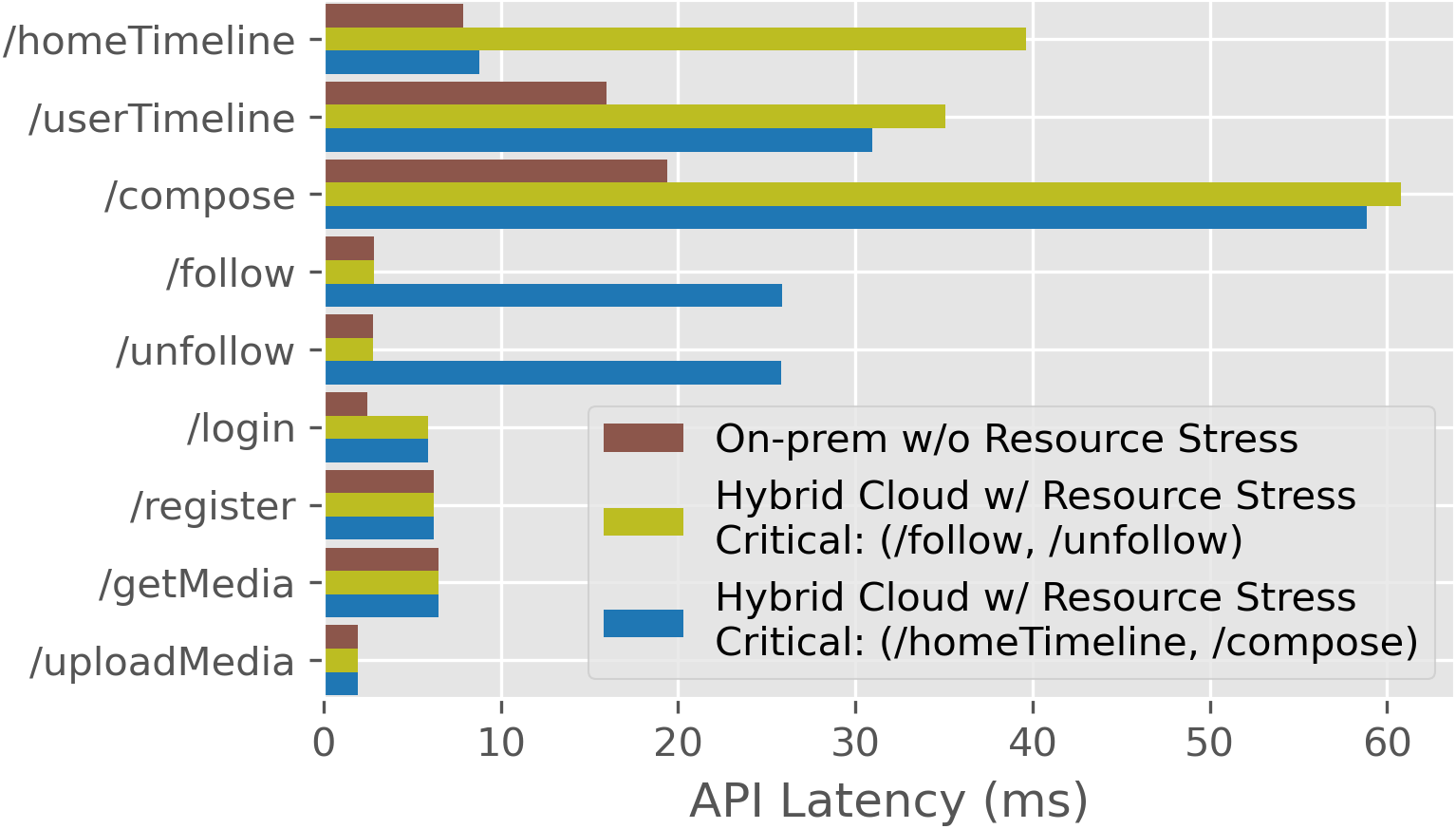}
	\caption{\scheme{} can personalize recommendations by considering the critical APIs specified by the application owner.}\label{fig:personalization}
\end{figure}

\scheme{} offers a unique feature for the application owner to specify critical APIs where responsive performance and high availability are crucial. We demonstrate such a capability with two example scenarios in Figure~\ref{fig:personalization}.
The first example (yellow bars) considers \texttt{/follow} and \texttt{/unfollow} to be critical. Compared with the API performance when all components were on-prem with no resource stress (brown bars), the performance-optimized plan by \scheme{} does not impact their responsiveness because the two critical APIs do not use the offloaded components for foreground operations. Considering the second scenario where \texttt{/homeTimeline} and \texttt{/compose} are critical (blue bars), the migration plan can no longer preserve the responsiveness of both \texttt{/follow} and \texttt{/unfollow} as they are not critical, and sacrificing them can create room for critical ones. Both \texttt{/homeTimeline} and \texttt{/compose} have an API latency shorter than the first example (yellow bars). Still, they suffer from performance impact because of the limited on-prem resources. Some components used by the critical APIs have to be moved, and long-distance communications are inevitable. 

\subsection{Post-migration Monitoring}\label{sec:exp-dyn}
\begin{figure}
	\centering
	\includegraphics[width=\linewidth]{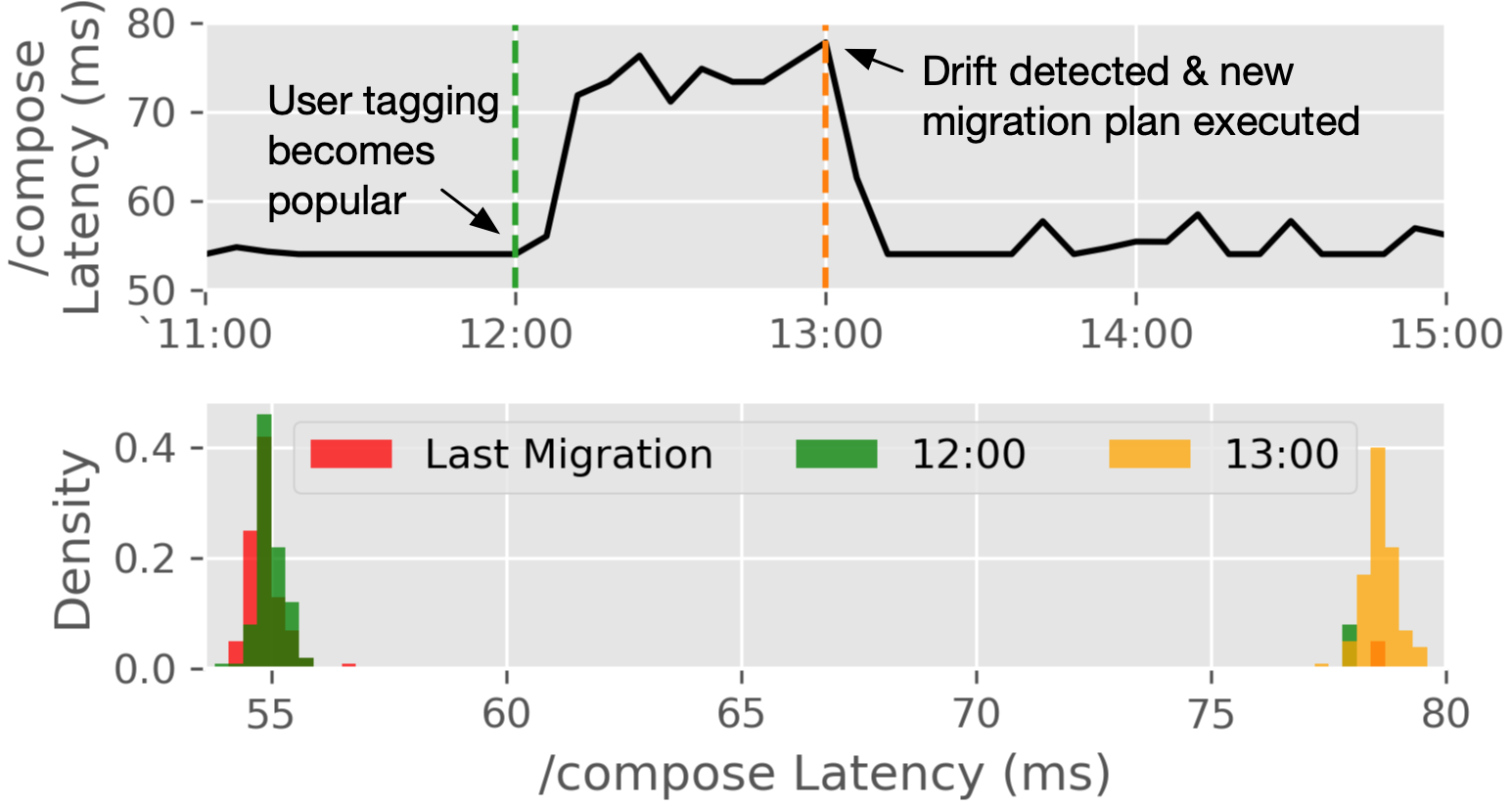}
	\caption{The new user behavior caused the latency of \texttt{/compose}  to increase (top). \scheme{} detects it according to the statistical discrepancy (bottom) and offers new plans to re-optimize the API latency.}\label{fig:dynamic-workload}
\end{figure}

\scheme{} offers proactive recommendations when better plans are available due to, e.g., changes in user behaviors. To evaluate it, we conduct an experiment with a dynamic workload in Figure~\ref{fig:dynamic-workload}, where users did not actively tag their friends in the social media posts before but have become active in doing so since 12:00. Such a change in user behaviors made the end-to-end latency of \texttt{/compose} to increase (see Figure~\ref{fig:dynamic-workload} (top)). This is because the migration plan executed in the previous round did not collocate \texttt{ComposePostService} and \texttt{UserMentionService}. The \texttt{/compose} requests under the new user behavior made these two components frequently interact, which were lengthy inter-datacenter communications. In this experiment, we run the post-migration monitoring module every hour, which is a hyperparameter configured by the application owner considering the frequency of migration recommendations. At 12:00, the latency distribution of \texttt{/compose} (see green bars in Figure~\ref{fig:dynamic-workload} (bottom)) still matched the one captured after the last migration (red bars). When \scheme{} conducts the same check with the latest requests at 13:00, the latency distribution (orange bars) shifts, leading to $29.74\times$ information loss in approximating the distribution (see Section~\ref{sec:monitoring}). \scheme{} begins searching for better plans, instructing the application owner to collocate \texttt{ComposePostService} and \texttt{UserMentionService}. With the updated plan, its end-to-end latency returns to the previous level under the new user behavior. 

\subsection{API Performance Modeling Analysis}\label{sec:exp-apiperf}
\begin{figure}
	\centering
	\includegraphics[width=0.6\linewidth]{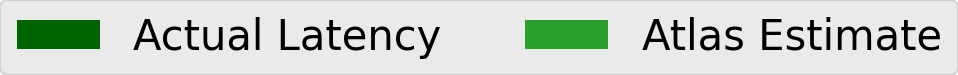}\\
	\begin{subfigure}[b]{0.49\linewidth}
		\centering
		\includegraphics[width=\linewidth]{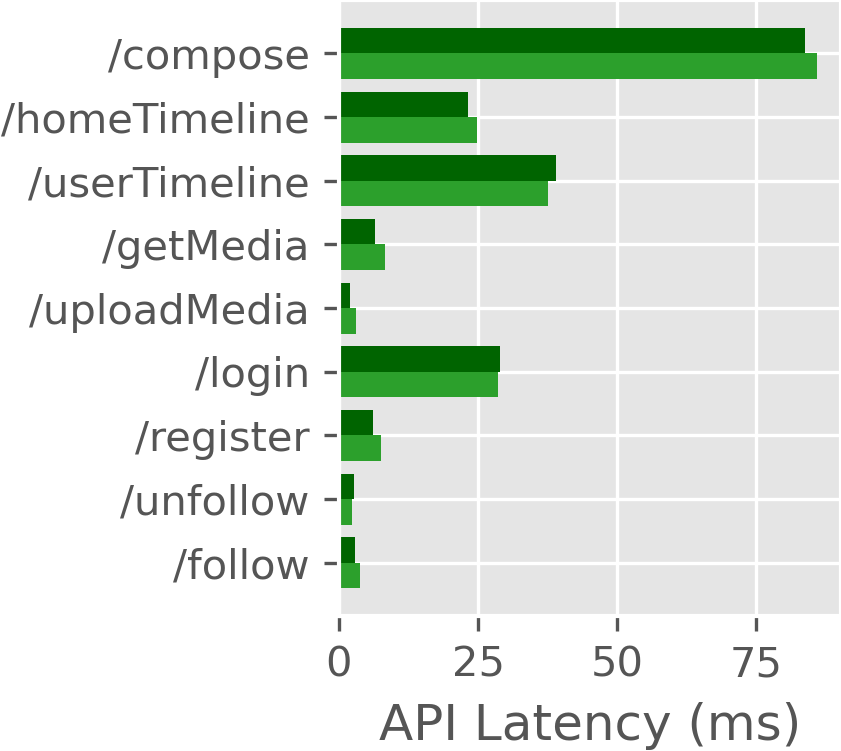}
		\caption{Performance-optimized Plan}
	\end{subfigure}
	\hfill
	\begin{subfigure}[b]{0.49\linewidth}
		\centering
		\includegraphics[width=\linewidth]{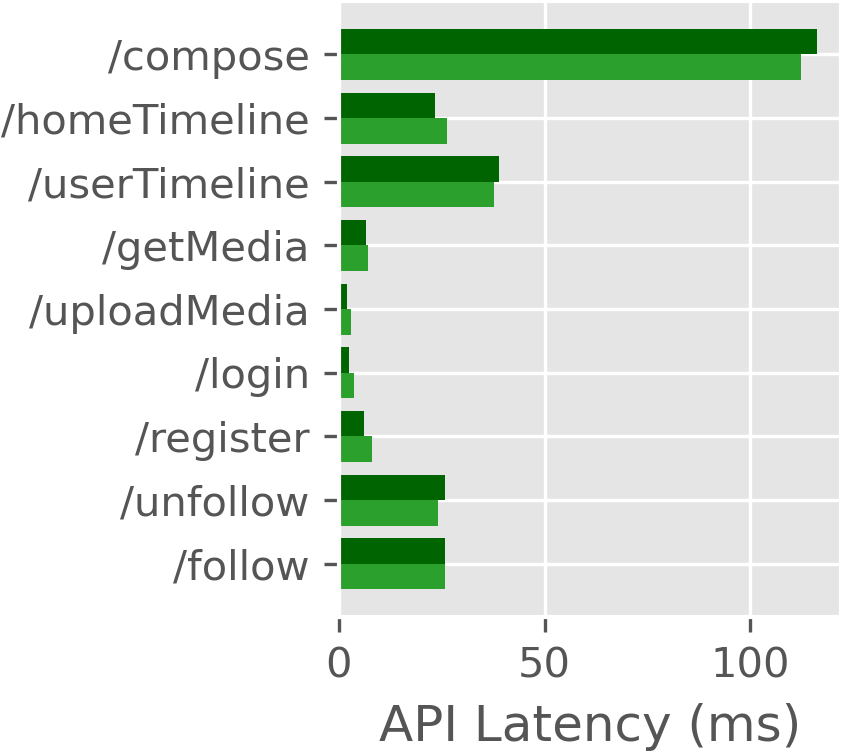}
		\caption{Cost-optimized Plan}
	\end{subfigure}
	\caption{\scheme{} can accurately estimate the API latency after migration without actual migration and measurements.}\label{fig:api-acc}
\end{figure}

\scheme{} gives application owners, for each recommended plan, a preview of API latency after migration. We analyze how accurate it is in Figure~\ref{fig:api-acc} with the performance-optimized and the cost-optimized plans as two examples. For each plan, we follow the suggested components to offload and record the API latency to compare with \scheme{}'s estimate. The preview (light green) offered by \scheme{} is accurate, with an error range of $3.97$ms. This means (i) the performance impact due to the deployment on hybrid clouds can be effectively characterized by our delay injection method, and (ii) the estimate serves as a useful reference for the application owner to understand how their application would perform after the migration.

\begin{figure}
	\centering
	\includegraphics[width=\linewidth]{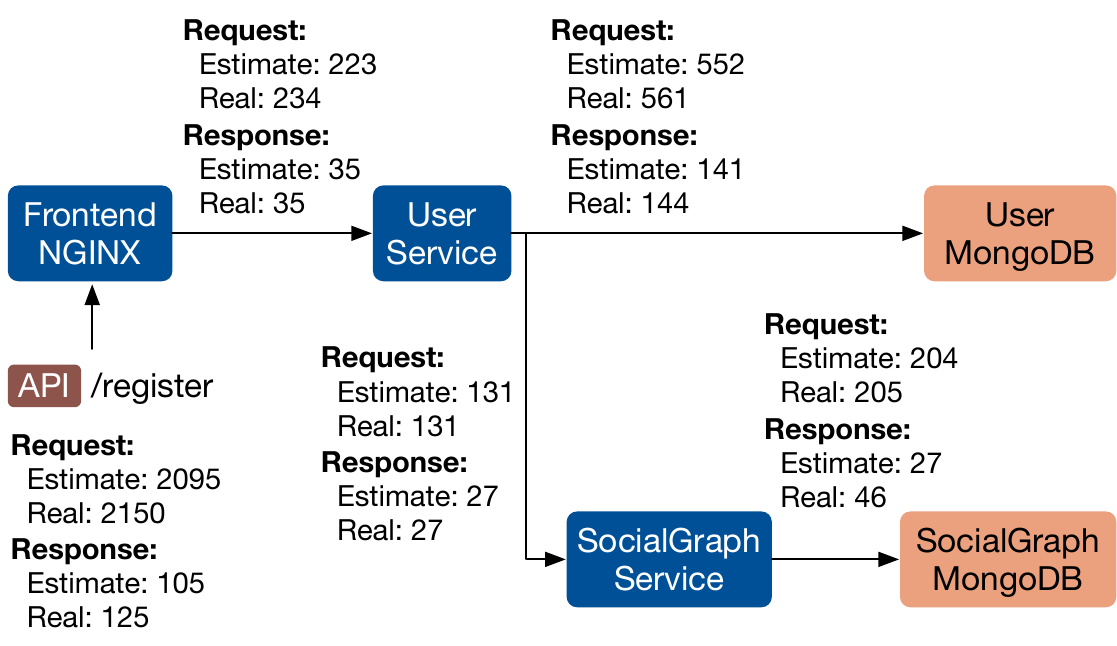}
	\caption{The learned network footprint of \texttt{/register} compared with the actual request and response sizes from a custom workload.}\label{fig:footprint-register}
\end{figure}

\begin{figure}
	\centering
	\includegraphics[width=\linewidth]{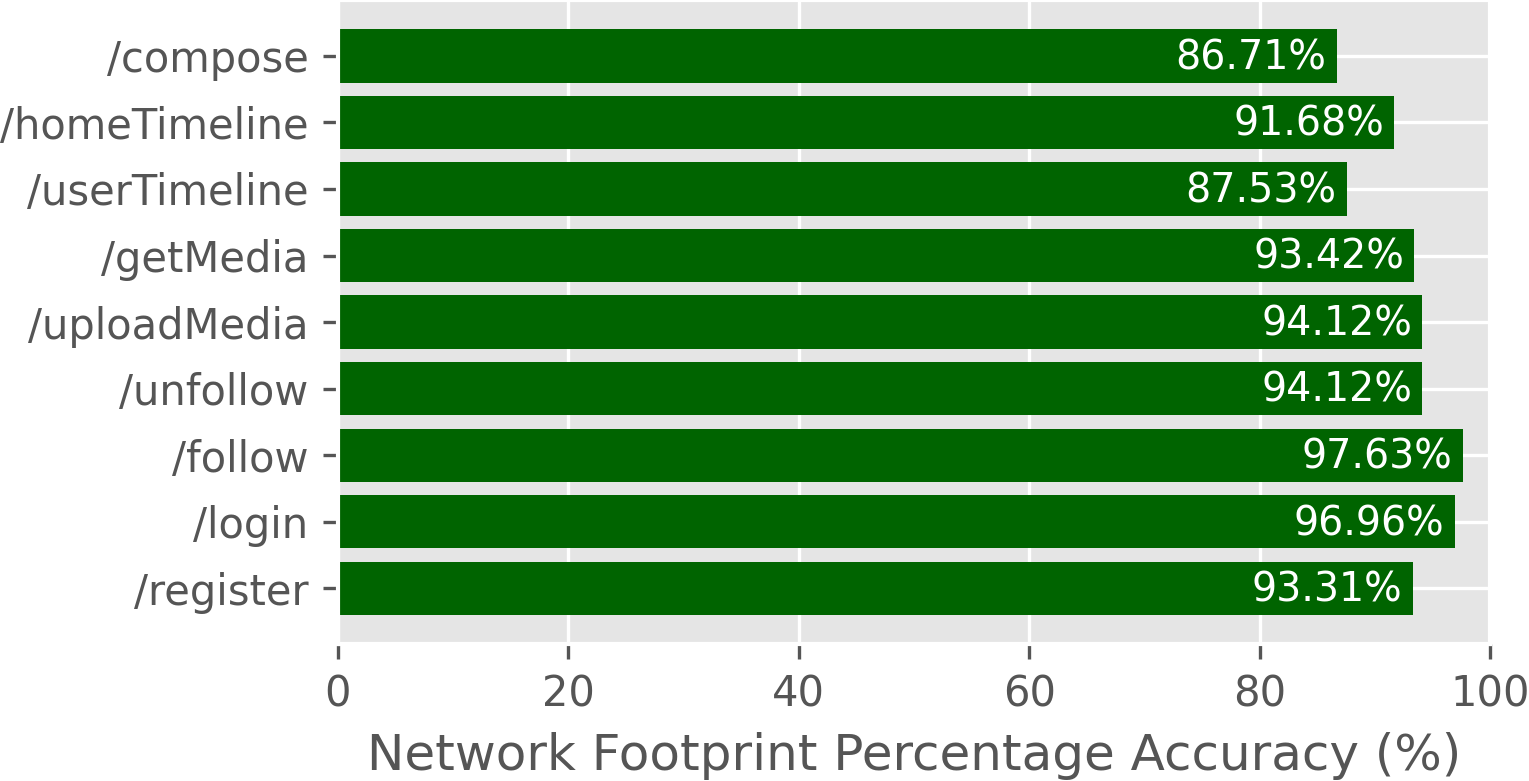}
	\caption{The network footprints learned by \scheme{} have high accuracy, matching the actual ones obtained by a custom workload.}\label{fig:footprint-all}
\end{figure}

A key contributing factor is the learning of network footprints. It helps inject an appropriate amount of delay into the API traces. We select an API with a moderate level of complexity, \texttt{/register}, for an in-depth study. Figure~\ref{fig:footprint-register} shows its learned network footprint compared with the real one captured by a custom workload. 
The network footprint of an API includes, for each communication from one component to another, how many bytes, on average, will be sent during the request and during the response. As shown in Figure~\ref{fig:footprint-register}, the estimate of data transfer size is similar to the real one. For example, the communication from \texttt{UserService} to \texttt{UserMongoDB} has an average request size of $561$ bytes and a response size of $144$ bytes. The learned footprint indicates an estimate of $552$ bytes and $141$ bytes, respectively. We summarize in Figure~\ref{fig:footprint-all} the percentage accuracy of network footprints of all nine APIs in the social network. 
It shows that \scheme{} can learn the footprints with accuracy from $86.71\%$ to $97.63\%$. In fact, the footprints are not just for API latency estimation but a multi-purpose by-product. We discuss how to use it to detect cyberattacks in Section~\ref{sec:discussion}.
\begin{figure}
	\centering
	\begin{subfigure}[b]{0.49\linewidth}
		\centering
		\includegraphics[width=\linewidth]{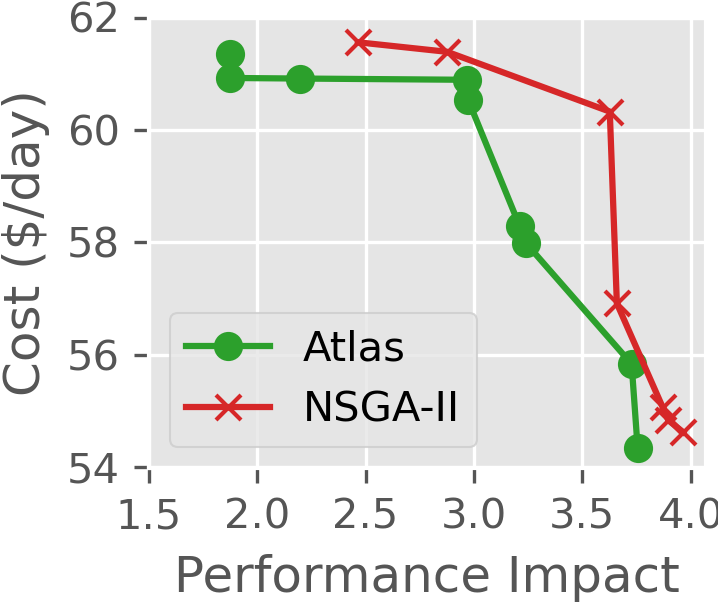}
		\caption{Pareto Front}\label{fig:rewd-driven}
	\end{subfigure}
	\hfill
	\begin{subfigure}[b]{0.49\linewidth}
		\centering
		\includegraphics[width=\linewidth]{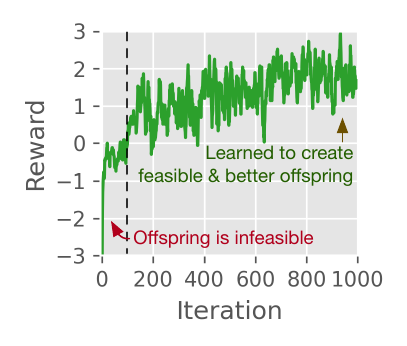}
		\caption{Reward Progression}\label{fig:rewd}
	\end{subfigure}
	\caption{Effectiveness of the DRL-based GA in \scheme{}.}
\end{figure}

\subsection{DRL-based Genetic Algorithm}\label{sec:exp-ablation}
\scheme{} features a reinforcement learning approach to formulate the crossover process in genetic algorithms as a reward-driven learning problem. We compare the Pareto front of \scheme{} ($9$ plans) with the one generated by a variant using NSGA-II as the GA ($7$ plans) in Figure~\ref{fig:rewd-driven}. They use the same quality modeling. For any plan offered by the NSGA-II approach, we can always find another plan along \scheme{}'s Pareto front with the same or, in most cases, better quality. To explain the effectiveness, we visualize the reward progression curve in Figure~\ref{fig:rewd}, where the RL model $\Lambda_{\boldsymbol{\theta}}$ was trained for $1,000$ iterations during the application learning phase. First, recall that a negative reward is given when the RL model cannot combine two parent plans into one that satisfies all constraints in Equation~\ref{eq:obj1} (e.g., exceeds on-prem limits). In the first $100$ iterations, the reward is consistently below zero, but it becomes positive afterward, which indicates that the RL model has learned to combine parent plans to create offspring that is at least deployable. This is important as \scheme{} does not waste time on those infeasible plans. Second, a higher reward reflects the RL model's capability to create offspring that outperforms its parents, which can be observed after the $100$th iteration. Therefore, \scheme{} can effectively find better plans out of millions. 

\section{Discussion}\label{sec:discussion}
\begin{figure}
	\centering
	\includegraphics[width=0.95\linewidth]{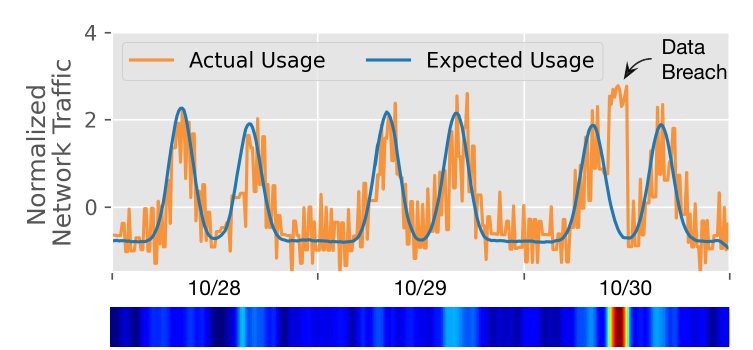}
	\caption{Using network footprints from \scheme{} to detect a data breach by attempting to justify the actual network traffic.}\label{fig:sanity}
\end{figure}

\textbf{Additional Use Case.} \scheme{}'s network footprint learning can be used to detect cyberattacks~\cite{chow2021sra}. The network footprint captures how many bytes are supposed to be sent and received between components when an API request is being served. By using the real API traffic specifying how many requests of different APIs have been made by the clients, one should be able to reconstruct the expected network traffic between all pairs of components, including unexpected workload spikes, which are also captured in the API traffic. This assists dedicated tools in detecting cybercriminals who infiltrate the application, e.g., through malicious Docker images~\cite{maliciousdocker} and copy data for ransom. Figure~\ref{fig:sanity} shows the network traffic from a MongoDB collected for three days (the orange line). By comparing it with the expected usage (the blue line) computed using the network footprints and the API traffic received within the same period (see the 1-D heatmap), we can identify a potential data breach on 10/30.

\textbf{Scalability.} \scheme{} is scalable to large applications. First, the observability tools have been widely used in production systems and incur minor overhead (e.g., $2.6\%$ on latency~\cite{gan2021sage} due to tracing). Many existing applications have already been installed with those tools for root-cause analysis. Second, the neural network for reward-based crossover can complete training in $19.21$ seconds and create offspring from the given parent plans in $0.459$ milliseconds. Increasing the number of components will enlarge the input dimension to the model, but its inference time grows sub-linearly due to the highly parallelizable computations (e.g., increasing the input dimensionality by $100\times$ is only $1.22\times$ slower). Finally, \scheme{} takes $34.2$ seconds to complete the recommendation. Genetic algorithms are known for their parallelizability, and the number of generations and population size can be increased accordingly to further widen the coverage~\cite{cantu1998survey,jin2008mrpga,harada2020parallel}. 

\textbf{Sky Computing.} Sky computing is a hot topic in the cloud industry~\cite{stoica2021cloud,sky-ibm}. It abstracts the complexity of multiple cloud vendors from the cloud users, who only need to specify high-level requirements. For example, SkyPilot~\cite{286502} has been introduced as a broker to provide computing infrastructure from multiple clouds for ML batch jobs. In microservices, several application components may only be assigned to specific clouds because of, e.g., privacy regulations about data placement~\cite{EUdataregulations2018} or the use of services unavailable elsewhere. We envision that \scheme{} can serve as a brain deciding how to distribute the remaining application components across clouds to optimize API latency, availability, and costs.
 
\section{Conclusions}
We have presented \scheme{}, a hybrid cloud migration advisor for interactive microservices. It considers user-facing APIs as first-class citizens, learns how each component is being used in the workflows of different APIs, and finds the best combination of components to offload, optimizing API latency, API availability, and cloud hosting cost. It is fueled by the telemetry data readily available in production systems without any human supervision. \scheme{} is an easy-to-deploy solution to guide the use of hybrid cloud toward elastic microservices. 

\section*{Acknowledgments}
We thank all reviewers for their insightful feedback. The first author acknowledges the IBM PhD Fellowship. The authors from the Georgia Institute of Technology are partially supported by the National Science Foundation under CISE Grants 2038029, 2302720, and 2312758, an IBM faculty award, and a grant from CISCO Edge AI program.

\bibliographystyle{ACM-Reference-Format}
\bibliography{reference}

\appendix
\section{Pricing Model}\label{app:pricing}
The pricing model used in our evaluation is generalized to reflect the key characteristics of different public clouds~\cite{aws,azure,google}. Public cloud providers often offer a list of node types. It is recommended to select the nodes with a similar hardware configuration for cluster autoscaler to operate~\cite{awsbpg}. 

\textbf{Compute.} Given the node type from the cloud provider with $\Omega^{\text{CPU}}$ cores and $\Omega^{\text{mem}}$ GB of memory, we can estimate the number of required nodes to serve the traffic at time $t$:
\begin{equation}
	n_t=\max\limits_{r\in\{\text{CPU, mem}\}}{\lceil(1+\delta^{r})\sum_{c\in\boldsymbol{\mathcal{C}}}\frac{p_c\tilde{\boldsymbol{\mathcal{U}}}^{r}_c[t]}{\Omega^{r}}\rceil},
\end{equation}
where $\delta^r$ controls the minimum free resources (e.g., $0.20$ to trigger scaling when the amount of free resource falls below $20\%$), $\boldsymbol{\mathcal{C}}$ is the set of components, and $\tilde{\boldsymbol{\mathcal{U}}}^{r}_c[t]$ is the expected usage of resource $r$ in component $c$ at time $t$ from the resource estimator (e.g., DeepRest~\cite{chow2022deeprest} used in this paper). Then, we obtain the compute cost
\begin{equation}
	\mathcal{Q}_{\text{compute}}^{\text{Cost}}(\boldsymbol{p})=\sum_t\Theta_{\text{compute}}n_t,
\end{equation}
where $\Theta_{\text{compute}}$ is the per-node price (e.g., $\$0.096$ for node type ``{m5.large}'' from AWS).

\textbf{Storage.} Autoscaling is not a privilege of compute but is also supported in cloud storage. We request $2\times$ the data size to be transferred during migration as the initial storage capacity $\kappa_0$, determine the capacity $\kappa_t$ at each time step $t$, and scale up whenever necessary~\cite{awsebs}:
\begin{equation}
	\hspace{-1.3em}\kappa_t=
	\begin{cases}
		\lceil(1+\delta^{\text{storage}})\kappa_{t-1}\rceil,& \text{if } 1-\frac{\sum_{c}p_c\tilde{\boldsymbol{\mathcal{U}}}^{\text{storage}}_c[t]}{\kappa_{t-1}}\leq\delta^{\text{storage}}\\
		\kappa_{t-1},              & \text{otherwise.}
	\end{cases}
\end{equation}
Then, we can get the storage cost 
\begin{equation}
	\mathcal{Q}_{\text{storage}}^{\text{Cost}}(\boldsymbol{p})=\sum_t\Theta_{\text{storage}}\kappa_t,
\end{equation}
where $\Theta_{\text{storage}}$ is the per-GB storage price (e.g., $\$0.08$ in AWS).

\textbf{Network Traffic.} Public clouds typically do not charge any data flow into their datacenters, but the egress traffic from the cloud can be expensive. Let $\Theta_{\text{traffic}}$ be the egress cost per GB (e.g., $\$0.09$ in AWS). The traffic-induced cost is:
\begin{equation}
	\hspace{-1.2em}\mathcal{Q}_{\text{traffic}}^{\text{Cost}}(\boldsymbol{p})
	=\sum_{t}\sum_{(c_i, c_j)\in\boldsymbol{\mathcal{C}}\times\boldsymbol{\mathcal{C}}}\mathbbm{I}[p_{c_i}\neq p_{c_j}]\Theta_{\text{traffic}}\tilde{\boldsymbol{\mathcal{U}}}^{\text{traffic}}_{c_i\rightarrow c_j}[t],
\end{equation}
where $\mathbbm{I}[\cdot]$ is a binary function returning $1$ if the condition is true and $0$ otherwise.

The overall cost of the plan $\boldsymbol{p}$ is 
\begin{equation}
	\mathcal{Q}^{\text{Cost}}(\boldsymbol{p})=\mathcal{Q}_{\text{compute}}^{\text{Cost}}(\boldsymbol{p})+\mathcal{Q}_{\text{storage}}^{\text{Cost}}(\boldsymbol{p})+\mathcal{Q}_{\text{traffic}}^{\text{Cost}}(\boldsymbol{p}),
\end{equation}
which has to be minimized.

\section{Artifact Appendix}\label{app:artifact}
\subsection{Abstract}
\scheme{} is a hybrid cloud migration advisor for interactive microservices. This artifact includes three components. To set up the experiment environment, we first provide a microservice-based social network application instrumented with distributed tracing and resource monitoring tools. We also provide an API traffic generator sending API requests with customizable workload characteristics. 
With the above two components, we provide the source code of \scheme{} to generate migration recommendations with a web-based interface for interactive analysis. The three components are released in a repository hosted on GitHub, and each is associated with a dedicated \texttt{README} file describing the setup and execution instructions.

\subsection{Description \& Requirements}
\subsubsection{How to access} The artifact is available on both GitHub and Zenodo.
\begin{itemize}[leftmargin=*]
	\item GitHub
	\begin{itemize}
		\item Link: \url{https://github.com/IBM/api-aware-cloud-migration}
		\item Hash: \textsc{7d7091c}
	\end{itemize}
	\item Zenodo
	\begin{itemize}
		\item Link: \url{https://doi.org/10.5281/zenodo.10080460}
	\end{itemize}
\end{itemize}

\subsubsection{Hardware dependencies}
\scheme{} has been tested on the following machine:
\begin{itemize}[leftmargin=*]
	\item Processor: Intel® Core i7-9700K CPU @ 3.60GHz × 8
	\item Graphics: GeForce RTX 2080 SUPER
	\item Memory: 32 GB
	\item Disk: 2.0 TB
\end{itemize}
The social network application and the traffic generator have been tested on the following nodes provided by CloudLab~\cite{Duplyakin+:ATC19}:
\begin{itemize}[leftmargin=*]
	\item On-premises: {c220g2}
	\item Cloud: {rs630}
\end{itemize}

\subsubsection{Software dependencies}
\scheme{} has been tested on Ubuntu 18.04.3 LTS and Python 3.7.

\subsubsection{Benchmarks} 
This artifact includes a benchmark and two datasets:
\begin{itemize}[leftmargin=*]
	\item Microservices: The social network application from DeathStarBench~\cite{gan2019open} with Jaeger\footnote{\url{https://www.jaegertracing.io}} for distributed tracing and Istio\footnote{\url{https://istio.io}} and Prometheus\footnote{\url{https://prometheus.io}} for resource monitoring.
	\item Social Network: The social graph~\cite{rossi2015network} for initialization.
	\item Media: The photos from INRIA Person~\cite{dalal2005histograms} for APIs related to media (e.g., \texttt{/uploadMedia}).
\end{itemize}

\subsection{Set-up}\label{sec:setup}
This subsection describes the preparation of the social network, the API traffic generator, and the migration advisor.

\subsubsection{Social Network Application}\label{sec:setup-sn} This artifact includes instructions to deploy the social network application with OpenEBS\footnote{\url{https://openebs.io}} as the storage engine. It can be launched using the YAML files provided in the  \texttt{social-network} directory. The \texttt{README.md} file provides step-by-step instructions.

\subsubsection{API Traffic Generator}\label{sec:api-traffic-generator}  We use Locust\footnote{\url{https://locust.io}} to implement the API traffic generator. The \texttt{locust} directory contains the source code with instructions provided in \texttt{README.md}. All required Python libraries can be installed with command:
\begin{verbatim}
	pip install -r requirements.txt
\end{verbatim}

\subsubsection{Hybrid Cloud Migration} 
The hybrid cloud migration advisor is implemented using Python and organized in the \texttt{migration-advisor} directory. It includes the scripts to process the data collected from the social network application and a web-based platform with precomputed simulations for demonstration. Similar to the API traffic generator, we provide all necessary Python libraries in \texttt{requirements.txt} and can be installed with the \texttt{pip} command as shown in Section~\ref{sec:api-traffic-generator}.

\subsection{Evaluation workflow}

\subsubsection{Major Claims}
Here are the major claims made in the paper:
\begin{itemize}[leftmargin=*]
	\item (C1): \scheme{} can provide migration recommendations with different trade-offs, namely API performance, API availability, and cloud hosting cost (E1). The results are illustrated in Figures 12-15.
	\item (C2): \scheme{} can offer personalized recommendations. The application owner can specify the budget, the API endpoints critical to their business, and the components that should not be migrated (E2). The results are illustrated in Figure 16.
	\item (C3): \scheme{} can conduct delay injection to accurately estimate the end-to-end latency of each user-facing API for a migration plan without actual execution (E3). The results are illustrated in Figures 18-20.
\end{itemize}

\subsubsection{Experiments}
Here are the experiments supporting the above major claims:\\

Experiment (E1): Hybrid Cloud Migration Recommendation [2 human-hour + 3 compute-hours]: 
This experiment generates API traffic to collect data for \scheme{} to learn about the application and requests migration recommendations to serve the expected API traffic that will consume more resources than the limit of the on-premises infrastructure. We can expect \scheme{} to provide multiple recommended plans with different trade-offs. Each option has a quality preview to help select the one that fits the business need.\\\\

\textit{[Preparation]} We first deploy the social network application following Section~\ref{sec:setup-sn} and obtain two addresses:
\begin{itemize}[leftmargin=*]
	\item \texttt{NGINX\_URL}: The address to the frontend NGINX server.
	\item \texttt{MEDIA\_URL}: The address to the media server.
\end{itemize} 
Then, we update the addresses in \texttt{locustfile.py} inside the \texttt{locust} directory and run the following command to load the social graph:
\begin{verbatim}
	python warmup.py --addr=NGINX_URL 
\end{verbatim} 

\noindent\textit{[Execution]} Follow the steps below to run this experiment:
\begin{enumerate}[leftmargin=*]
	\item Run \texttt{locust -f locustfile.py} to send 100 minutes of API requests. The first 30 minutes are used for application learning. The remaining 70 minutes have $3\times$ more user requests to simulate the expected API traffic. 
	\item After finishing the load generation, go to the \\\texttt{migration-advisor} directory.
	\item Follow the postprocessing instructions in \texttt{README.md} to extract traces and resource utilization.
	\item Follow the instructions in \texttt{README.md} to run \texttt{app.py} and launch the web-based interface either with the data collected in this experiment or the precomputed simulations.
	\item Go to the \texttt{Migration Dashboard} and click \texttt{Recommend}.
\end{enumerate}
~\\
\textit{[Results]} You will be given the recommended migration plans on a 3D scatter plot (a list view is also provided). The three dimensions correspond to API performance, API availability, and cloud hosting cost. You can click the dot, and the details of the corresponding plan will be provided on the right-hand side, including which components should be moved to the cloud, the cloud hosting cost, the number of APIs that will be unavailable during the migration process, and the end-to-end latency of each API before and after migration.
~\\

Experiment (E2): Personalized Recommendation [10 human-minutes + 10 compute-minutes]: 
This experiment uses the web-based interface to show that \scheme{} can personalize recommendations based on the preferences specified by the application owner. \\\\
\textit{[Preparation]} We follow the preparation for Experiment (E1) to deploy the social network and the web-based interface.\\
~\\
\textit{[Execution]} Follow the steps below to run this experiment:
\newpage
\begin{enumerate}[leftmargin=*]
	\item Specify different budgets and observe the cost of each recommended migration plan.
	\item Specify different API endpoints as critical and observe the estimated end-to-end latency of critical APIs of the performance-optimized plan.
	\item Specify different components as on-prem placement constraints and observe the on-prem components recommended by each plan.\\
\end{enumerate}

\noindent\textit{[Results]} You can observe that \scheme{} will only provide the migration plans below the budget if it is given as a preference. For critical API endpoints, their end-to-end latency will tend to be lower than the preferences that consider them to be non-critical. Finally, when on-prem placement constraints are provided, \scheme{} will ensure that those components will not be migrated to the cloud. 	~\\

Experiment (E3): API Latency Estimation [30 human-minutes + 2 compute-hour]: 
This experiment shows the accurate estimation of end-to-end latency with delay injection. \\\\
\textit{[Preparation]} We follow the preparation for Experiment (E1) to deploy the social network and the web-based interface.\\
~\\
\textit{[Execution]} Follow the steps below to run this experiment:
\begin{enumerate}[leftmargin=*]
	\item Go to the \texttt{Migration Dashboard}, click \texttt{Recommend}, and record the performance-optimized plan.
	\item Restart the social network application and run the API traffic with the locust script.
	\item At the 30th minute, follow the performance-optimized plan to offload components.
	\item Visit the Locust webpage and observe the end-to-end latency of different API endpoints.\\
\end{enumerate}

\noindent\textit{[Results]} You can observe that  the estimated latency of an API provided in the quality preview of a migration plan is close to the actual latency measured after following the plan for migration. 

\end{document}